\documentclass[preprint]{JHEP3} % 10pt is ignored!
\JHEPspecialurl{http://jhep.sissa.it/JOURNAL/JHEP3.tar.gz}
\usepackage{amssymb,epsfig,multicol}
\newcommand\fverb{\setbox\fverbbox=\hbox\bgroup\verb}
\newcommand\fverbdo{\egroup\medskip\noindent%
\fbox{\unhbox\fverbbox}\ }
\newcommand\fverbit{\egroup\item[\fbox{\unhbox\fverbbox}]}
\newbox\fverbbox

\catcode`@=11
\def\secteqno{\@addtoreset{equation}{section}
\def\theequation{\thesection.\arabic{equation}}}
\catcode`@=12
\secteqno
\newcommand{\be}{\begin{equation}}
\newcommand{\ee}{\end{equation}}
\newcommand{\bea}{\begin{eqnarray}}
\newcommand{\eea}{\end{eqnarray}}
\newcommand{\bref}[1]{(\ref{#1})}
\newcommand{\nn}{\nonumber}
\newcommand{\A}{\alpha} \newcommand{\B}{\beta} \newcommand{\gam}{\gamma}
\newcommand{\G}{\gamma} \newcommand{\D}{\delta} 
\newcommand{\ep}{\epsilon} 
\newcommand{\T}{\theta} 

\newcommand{\lam}{\lambda}
\newcommand{\s}{\sigma}\newcommand{\Gam}{\Gamma}

\newcommand{\h}{\eta}           
           
\newcommand{\W}{\Omega}         
            \newcommand{\Sig}{\Sigma}
 
\newcommand{\ov}[1]{\overline #1}

\newcommand{\ba}{\overline }
\def\6{\partial}
\def\7{\tilde}
\def\8{\hat}

\def\CC{{\cal C}}\def\CG{{\cal G}}\def\CL{{\cal L}}

\def\CQ{{\cal Q}}\def\CB{{\cal B}}\def\CD{{\cal D}}
   \def\CM{{\cal M}} \def\CP{{\cal P}}
\def\CM{{\cal M}}\def\CJ{{\cal J}}
\def\CB{{\cal B}}

\def\vs{\vskip 4mm}\def\={{\;=\;}}\def\+{{\;+\;}}
\def\bL{{\rm{\bf L}}}
\def\bL{{\rm{\bf L}}}\def\rB{{{B}}}
\def\bQ{{\rm{\bf Q}}}\def\bS{{\bf {\Sigma}}}
\def\OSp{{$OSp(2|4)$ }}\def\usp{{$UU_\alpha (1,1|1;H)$}}
\def\SO{O}%\def\SO{SO}
\def\too{{\;\to\;}}\def\vs{\vskip 4mm}
\newcommand{\sM}{Maxwell superalgebra}\newcommand{\sMs}{Maxwell superalgebras}
\newcommand{\Jis}{Jacobi identities }

\newcommand{\braket}[2]{\langle #1|#2\rangle}
\title{ Deformations of Maxwell Superalgebras and Their Applications}

\author{{Sotirios Bonanos}$^1$,~
{Joaquim Gomis}$^2$,~{Kiyoshi Kamimura}$^3$ and {Jerzy Lukierski}$^4$ \\
{${}^1$~Institute of Nuclear Physics, NCSR Demokritos,
15310 Aghia Paraskevi, Attiki, Greece}\\
{${}^2$~Departament ECM and ICCUB, Universitat de Barcelona,
Diagonal 647, 08028 Barcelona, Spain } \\
{${}^3$~Department of Physics, Toho University,
Funabashi, 274-8510 Japan} \\
{${}^4$~Institute of Theoretical
Physics, Wroclaw University, pl. Maxa Borna 9, 50-204 Wroclaw, Poland} \\
}
\received{\today}       %%
\accepted{\today}       %% These are for published papers.

\preprint{ICCUB-10-032, Toho-CP-1093} 
\abstract{
We describe the Lie algebra deformations of D=4 Maxwell superalgebra that was recently introduced as the symmetry algebra of a kappa-symmetric massless superparticle in a supersymmetric constant electromagnetic background.  Further  we introduce the D=3 Maxwell superalgebra and present all its possible deformations. Finally the deformed superalgebras are used to derive via  a contraction procedure the complete set of Casimir operators for D=4 and D=3 Maxwell superalgebras.
\vs
}

\begin{document}

%\maketitle  IS IGNORED %%%%%%%%%%%

\section{Introduction}
The Poincar{\`e} algebra and Poincar{\`e} group describe the symmetries of empty Minkowski space-time. Filling such a flat space-time  with some background  fields leads to a modification of Poincar{\`e} symmetries.  An example of such a modification  is the so-called Maxwell symmetries, which was obtained already in the seventies \cite{Bacry:1970ye}\cite{Schrader:1972zd} by considering Minkowski space with an added constant electromagnetic (EM) background. The collection of arbitrary values of the constant EM field strengths provides additional degrees of freedom in Minkowski space, supplementing  the Poincar{\`e} group with additional group parameters and the Poincar{\`e} algebra with new generators.

The Maxwell algebra \cite{Schrader:1972zd}-\cite{Galperin:1987wb},
see also \cite{Cangemi:1992ri}-\cite{Duval:2008tr} for D=3, is obtained by adding to the Poincar{\`e} generators $(P_\mu,\,M_{\mu\nu})$ the tensorial central charges $Z_{\mu\nu}$ ($Z_{\mu\nu}=-Z_{\nu\mu}$) which modify the commutativity of the four-momenta $P_\mu$
\be\label{max}
 [P_\mu,\,P_\nu]=i\,Z_{\mu\nu},
\ee
where $M_{\mu\nu}$ are the Lorentz algebra generators and
\bea
\left[Z_{\mu\nu},M_{\rho\s}\right] &=& -i\,\h_{\nu[\rho}Z_{|\mu|\s]}+i\,
\h_{\mu[\rho}Z_{|\nu|\s]}, \\
\left[P_\mu,\, Z_{\rho\s}\right] &=& [Z_{\mu\nu},\,Z_{\rho\s}]=0.
\label{max1}\eea
The D-dimensional Maxwell algebra $G=(M_{\mu\nu},\,P_\mu,\,Z_{\mu\nu})$ has the
structure of a semi-direct sum
\be
{G}=\SO(D-1,1) {+}\hskip-2.8mm{\supset} H  
\ee
where the algebra $H$ $(=(P_\mu,\,Z_{\mu\nu})$) can be obtained by suitable contraction $\alpha \to 0$ of the de Sitter algebra $\SO(D,1)=({\cal M_{\mu\nu}},\,{\cal P_\mu})$,  or anti-de Sitter $\SO(D-1,2)$, with \cite{Bonanos:2009wy}
\be
{\cal M_{\mu\nu}}=\frac{1}{\alpha^2}Z_{\mu\nu}, \quad {\cal P_\mu}=\frac{1}{\alpha}P_\mu.
\ee

The Maxwell algebras and Maxwell symmetries were recently studied in three different directions:
\begin{enumerate}
\item
The Maxwell algebra is an enlargement of Poincar{\`e} algebra, i.e. by putting $Z_{\mu\nu}=0$ one gets back to the  Poincar{\`e} algebra. Analogously, one can consider  the class of supersymmetrizations providing \sM \ as the minimal enlargement of N=1 Poincar{\`e} superalgebra. Such supersymmetric extension of D=4 Maxwell algebra was obtained in \cite{Bonanos:2009wy}  by adding minimal number of two four-dimensional Majorana supercharges $\bQ_\alpha,\bS_\alpha$ and  mathematically optional two scalar generators $B_5,\,B$. The coset $\frac {SuperMaxwell}{Lorentz\times B_5}$
describes the supersymmetries of flat (Wess-Zumino) Minkowski superspace with  arbitrary constant values of an Abelian gauge superfield background
\be
 \label{eqjap1.4}
 W_\alpha (\theta) = i\,\lambda_\alpha -
 \frac{i}{2} \, f_{\mu\nu} (\bar{\theta} \gamma^{\mu\nu})_\alpha
 - i D (\bar{\theta} \gamma_5)_\alpha.
\ee
The superspace coordinates  $(x^\mu,\,\theta^\alpha,\phi)$ are supplemented in the framework of Maxwell supergeometry by graded additional coordinates $(\lambda_\alpha,f_{\mu\nu},D)$ related to the generators  $(\bS_\alpha,Z_{\mu\nu},B)$.

\item Following preliminary results obtained in \cite{Soroka:2004fj}, all deformations of the Maxwell algebra in any dimension D=d+1 were studied recently in \cite{Gomis:2009vm}.  In arbitrary dimension D there is a ``universal" $k$-deformation,   resulting in the following deformed Maxwell algebras:
\bea
k>0&:&  \SO(d,1)\oplus \SO(d,2)\qquad  (Lorentz \oplus AdS)\nn \\
k<0&:&  \SO(d,1)\oplus \SO(d+1,1)\quad (Lorentz \oplus dS) \label{kdeforfStruct}
\eea
 In D=3 the deformations are parametrized by two parameters $(k,\,b)$,
with an additional ``exotic" $b$-deformation.
The $(k,\,b)$ plane can be divided into two domains where the two deformed Maxwell algebras described by \bref{kdeforfStruct} are realized. However on the curve separating these two domains the obtained algebra is isomorphic to $\SO(2,1)\oplus ISO(2,1)$ (D=3 Lorentz $\oplus$ D=3 Poincar{\`e}).
\item{
One can study further extensions of the Poincar{\`e} symmetries by adding new tensorial central generators \cite{Bonanos:2008kr}\cite{Bonanos:2008ez}  to the  first level extension described by the Maxwell algebra. For example, in D=4 the second level extension consists in adding the third  rank tensorial charges $Y_{\mu[\rho\s]}$, which can be related to a Minkowski space filled with arbitrary linear EM background (i.e., $F_{[\mu\nu]}=f_{[\mu\nu]}+f_{[\mu\nu]\rho} \,x^\rho$, $f_{[\mu\nu]}$ and $f_{[\mu\nu]\rho}$ arbitrary constant tensors). }
\end{enumerate}
\vs

The  first aim of this paper is to consider all possible deformations of the D=4 Maxwell superalgebra
introduced  in   \cite{Bonanos:2009wy}.
In D=4 one obtains two independent deformations:
\begin{itemize}
\item First, the supersymmetrization of the ``universal" $k$-deformation given by \bref{kdeforfStruct}. Because of the doubling of Majorana supercharges in the  \sM, the deformed superalgebras in D=4  require also eight real  supercharges describing N=2 AdS and  N=1 dS SUSY\footnote{For the simplest N=1 supersymmetrization of D=4 de-Sitter algebra we need 8 supercharges (see Appendix B)}
\bea
k>0&:&  \SO(3,1)\oplus OSp (2|4) \oplus R \label{susykdeforfStruct1}\\
k<0&:&  \SO(3,1)\oplus UU_\alpha (1,1|1;H) \oplus R \label{susykdeforfStruct2}
\eea
where \usp \ is N=1, D=4 de-Sitter superalgebra \cite{Kugo:1982bn}\cite{Lukierski:1984it}\cite{Pilch:1984aw} and will be explained in detail in Appendix B.
\item  Second, the $s$-deformation which does not have a non-SUSY counterpart. It involves only a modification of the algebraic relations for the scalar generator $B_5$ with the dilatation operator $\CD$ given by the replacement $B_5\to B_5+s\,\CD$. If we enlarge   the Maxwell superalgebra
by Weyl symmetry then the $s$-deformation is no longer an independent deformation\footnote{An analogous situation appears in the deformations of the symmetries of very special relativity \cite{Gibbons:2007iu}.}.
\end{itemize}

Second aim of the paper is to introduce the Maxwell superalgebra in D=3 and study its possible deformations. The additional $s$-deformation is not present in D=3. We find that there is a two parameter deformation of D=3 \sM \, as in the bosonic case considered in \cite{Gomis:2009vm}.
Depending on the values of the deformation parameters, we find three different deformed superalgebras: $\SO(2,1)\oplus OSp(1|2)\oplus OSp(1|2)$,
$\SO(2,1)\oplus  OSp(1|2;C)\oplus{\overline{OSp(1|2;C)}}$ and $OSp(1|2)\oplus$(D=3 superPoincar{\`e}).
\vs

The plan of the paper is the following: In section 2 we recall the results on D=4 Maxwell superalgebra \cite{Bonanos:2009wy} and introduce its one-dimensional Weyl extension by adding   appropriate  scale transformations\footnote{ One could also consider the enlargement of pure bosonic Maxwell algebra by dilatations.}. It appears that the superMaxwell-invariant massless superparticle model, introduced in  \cite{Bonanos:2009wy}, is also invariant under the Maxwell-Weyl  supersymmetry.
In section 3 we discuss two deformations of the D=4 Maxwell superalgebra.
One ($k$-deformation)  is described by the superalgebras \bref{susykdeforfStruct1} and \bref{susykdeforfStruct2},
and the other ($s$-deformation) can be  introduced as a parameter-dependent class of subalgebras of the D=4 Maxwell-Weyl superalgebra.
In section 4 we introduce the D=3 Maxwell superalgebra and obtain the supersymmetrization of the two-parameter family of deformations.

In  \cite{Bonanos:2009wy} we have presented the
bilinear Casimir operators of the D=4 Maxwell superalgebra, including the generalized mass-shell formula.
In section 5, by contracting the known Casimir operators of \bref{susykdeforfStruct1} ($k \to 0$) we obtain all   six Casimirs of the D=4 Maxwell superalgebra.
Subsequently, the Casimir operators for D=3 \sM \ are also obtained  via contraction.
In section 6 we present conjectures about the existence of Maxwell superalgebras for
$D>4$ and its deformations, outline the relation with other proposals  \cite{Bergshoeff:1995hm}\cite{Soroka:2010ht} and conclude with some final remarks. We add also two appendices, one summarizing our conventions including gamma matrices, and a second describing quaternionic (super) groups and (super) algebras as well as the D=4, N=1 de-Sitter superalgebra \usp \ appearing in \bref{susykdeforfStruct2}.

%%%%%%%%%%%%%%%%%%%%%%%%%%%%%%%%%%%%%%%%%%%%%%%

\section{D=4 \sM\,  and its Weyl-enlargement}
 In a recent paper \cite{Bonanos:2009wy} we have proposed the following supersymmetric
extension, denoted by ${\cal G}_5$ , of the Maxwell algebra in 4 dimensions
(our notations and conventions are summarized in Appendix A),
\bea
\left[P_\mu,P_{\nu}\right]&=&i\,Z_{\mu\nu},\qquad\qquad
\left[P_\mu,{\bQ}_\A\right]=-i\,\bS_{\B}(\G_{\mu}{)^{\B}}_\A,
\nn\\
\{{\bQ}_\A,{\bQ}_\B\}&=&2\,(C\G^{\mu})_{\A\B}P_\mu,\qquad
\{{\bQ}_\A,{\bf\Sigma}_{\B}\}=\frac12(C\G^{\mu\nu})_{\A\B}\,Z_{\mu\nu}\,+\,
(C\G_5)_{\A\B}\,{\rB},
\nn\\
\left[B_5,{\bQ}_\A\right]&=&-i\,(\bQ\G_{5})_\A,\qquad
\left[B_5,{\bS}_\A\right]=\,i\,(\bS\G_{5})_\A,
\nn\\
\left[P_\mu,M_{\rho\s}\right]&=&-i\,\h_{\mu[\rho}P_{\s]},
\qquad
\left[Z_{\mu\nu},M_{\rho\s}\right]=-i\,\h_{\nu[\rho}Z_{|\mu |\s]}+i\,
\h_{\mu[\rho}Z_{|\nu |\s]},
\nn\\
\left[M_{\rho\s},\bQ_{\A}\right]&=&-\frac{i}2(\bQ \G_{\rho\s})_{\A},\qquad
\left[M_{\rho\s},\bS_{\A}\right]=-\frac{i}2(\bS \G_{\rho\s})_{\A},
\nn\\
\left[M_{\mu\nu},M_{\rho\s}\right]&=&-i\,\h_{\nu[\rho}M_{| \mu |\s]}+i\,
\h_{\mu[\rho}M_{|\nu |\s]}.
\label{newalgebra}\eea

The bosonic generators  $(P_\mu,M_{\mu\nu},Z_{\mu\nu})$, linked to translations,
Lorentz rotations and additional tensorial coordinates, form the bosonic Maxwell subalgebra and the fermionic generators
$\bQ_\A,{\bS}_\A,\,(\A=1,2,3,4)$ are two Majorana spinor charges.
$B$ is a central charge and $B_5$ generates chiral transformations. We point out that D=4  Maxwell superalgebra can also  be considered as an enlargement by generators $Z_{\mu\nu}$ of the algebra with 8 supercharges introduced by Green \cite{Green:1989nn}.

 There are three subalgebras obtained by  consistently removing generators $B$ and/or $B_5$ from \bref{newalgebra} (see \cite{Bonanos:2009wy}).

1) The minimal supersymmetric extension $\CG$, with a bosonic sector consisting only of the Maxwell algebra generators, is obtained  if we remove $B$ and $B_5$.

2) Removing only generator $ B_5$, we get a central extension $\7\CG$ of $\CG$. The generator $B$ is required if we wish to introduce the scalar degree of freedom
describing the off-shell extension of D=4 $U(1)$ field strength supermultiplet.

3) One can consider a subalgebra with only  the generator $B_5$ which acts on the supercharges $\bQ_{\A},\,\bS_{\A}$ as chiral generator. If $B$ is present,  $B_5$ is also required for the existence of the supersymmetric mass Casimir.

In this paper we shall consider the Maxwell superalgebra $\CG_5$ with  both $B $ and $B_5 $ given in \bref{newalgebra}.
We add that all cases describe the supersymmetric extension of the Maxwell algebra with minimal number of supercharges (eight real or four complex) and all these supersymmetrizations describe N=1 \sM \footnote{We mention that the superalgebra $\SO(3,1)\oplus OSp(1|4)$ with four real supercharges considered in \cite{Soroka:2006aj}\cite{Soroka:2010ht} describe the supersymmetrization of one of the deformations of \sM \ (see also section 6).}. We note that four  additional supercharges $\bS_\alpha$ are  present due to  the supersymmetrization of  the constant electromagnetic background
\bref{eqjap1.4}.

The superalgebra $\CG_5$
describes the symmetries of the massless  kappa invariant  superparticle action  in an external  constant N=1 susy invariant background presented in \cite{Bonanos:2009wy}
\be
\CL=\frac{\pi_\mu^{2}}{2e}+\frac12f_{\mu\nu} {L}_Z^{\mu\nu}+{{i}}\lambda_\A  {L}_\Sig^\A +D  {L}_B, \label{Lag0}
\ee
where
$ {L}_P^\mu=\pi^\mu=dx^\mu+i\bar\T\G^\mu\dot\T, \; {L}_Z^{{\mu\nu}}, \; {L}_\Sig^\A,\,  {L}_B$ are the pullbacks on the world line of the components of the MC forms $ {\Omega} = -i  {g}^{-1} d  {g}$ defined on the supercoset ${{\CG}_5}/({Lorentz}\otimes B_5)$
\be
 {g}= e^{\frac{i}2Z_{\mu\nu}\phi^{\mu\nu}}\,
e^{iP_\mu x^\mu}\,  e^{i\bS_{\A}{\phi}^{\A}}\, e^{{i\bQ}_\A\T^\A}\,e^{i{B}\,\phi}.
\label{coset}\ee

We observe that one can assign
mass dimensions to the generators of the superalgebra \bref{newalgebra} as follows
\be
[P_\mu]=1,\quad[Z_{\mu\nu}]=2,\quad[\bQ_\A]=\frac12,\quad[\bS_\A]=\frac32,\quad[B]=2, \quad[B_5]=[M_{\mu\nu}]=0,\ee
which can be
described by introducing a dilatation generator $\CD$ satisfying the relations:
\bea
\left[\CD,\, P_\mu\right]&=& i\,P_\mu, \quad [\CD,\, Z_{\mu\nu}]=2\, i\,Z_{\mu\nu},\quad [\CD,\, \bQ_{\A}]=
\frac{i}{2}\,\bQ_{\A},\nn \\
\left[\CD,\, \bS_{\A}\right]&=& \frac{3}{2}\,i\,\bS_{\A},\quad[\CD,\,B]=2i, \quad [\CD,\,B_5]=[\CD,\,M_{\mu\nu}]=0.
\label{Daction}
\eea
The supercoset coordinates in \bref{coset} transform under the scale transformations, generated by $\CD$, with opposite mass dimensionalities:
\be
x^{\mu '} = \lambda ^{-1} x^\mu,\quad
  {\phi^{\mu\nu '}} = \lambda ^{-2} \phi^{\mu\nu},\quad
\theta^{\alpha '}= \lambda ^{-1/2} \theta^\alpha, \quad
 \phi^{\alpha '}= \lambda ^{-3/2} \phi^{\alpha}, \quad
\phi' =\lambda ^{-2}  \phi. \label{coordScal}
\ee

 Adding relations \bref{Daction} to \bref{newalgebra} one obtains the Maxwell-Weyl superalgebra, which is a one-dimensional  enlargement of Maxwell superalgebra described by the semidirect sum $\CD {+}\hskip-3mm{\supset}{\CG}_5$.
We note that the massless superparticle action \bref{Lag0} remains invariant under the scale transformation with the einbein, transforming as
 $e'=\lambda^{-2}e$, consistent with its  role as a coordinate for $\CD$
in the coset ${\CD {+}\hskip-3mm{\supset}{\CG}_5}/({Lorentz}\otimes B_5)$.
\vs

%%%%%%%%%%%%%%%%%%%%%%%%%%%%%%%%%%%%%%%%%%%%%
\section{Deformations of \sM \ in D=4}

The \sM \  \bref{newalgebra} is equivalently described in terms of the Maurer Cartan
one form
\bea
\W&=&P_\mu L_P^\mu+\frac12M_{\mu\nu}L_M^{\mu\nu} +\frac12 Z_{\mu\nu} L_Z^{\mu\nu}+ B L_B +B_5L^5 +\bQ_\A\bL^\A+\bS_\A\bL_\Sig^\A,
\label{newalgebraMC}\eea satisfying the Maurer Cartan
equation $d\W+i\W\wedge \W=0$, as
\bea
 d {L}_P^\mu+ {L}_M^{\mu\nu} {L}_{P\nu}-{i}{\overline { {\bL}}}\G^{\mu} {\bL}
&=& 0,
\nn\\  d {L}_M^{\mu\nu}+  {L}_M^{\mu\rho}\h_{\rho\s} {L}_M^{\s\nu}&=&0,
\nn\\
 d{L}_Z^{\mu\nu}+ {L}_M^{\mu\rho}\h_{\rho\s}{L}_Z^{\s\nu}
+{L}_Z^{\mu\rho}\h_{\rho\s}{L}_M^{\s\nu}- {L}_P^\mu\, {L}_P^\nu
- {i}{\overline {  {\bL}}}\G^{\mu\nu}  {\bL}_\Sig
&=&0,
\nn\\
 d {\bL}^\A+\,\frac14{L}_M^{\mu\nu}(\G_{\mu\nu} {\bL})^\A
+\,  {L}^5\,(\G_5  {\bL})^\A\,
&=&0 ,
\nn\\
d {\bL}_\Sig^{\A}+\frac14 {L}_M^{\mu\nu}(\G_{\mu\nu} {\bL}_\Sig)^\A
+\,  {L}_P^\mu(\G_{\mu}{\bL})^{\A}-\, {L}^5(\G_5  {\bL}_\Sig)^\A
&=&0,
\nn\\
d {L}_B- {i}{\overline { {\bL}}}\,\G_{5}\, {\bL}_\Sig&=&0,
\nn\\
 d {L}^5&=&0.
\label{MCBS}\eea
These MC equations provide a dual formulation  of the \sM \  \bref{newalgebra}
and closure of the system \bref{MCBS} under exterior differentiation  is equivalent to the Jacobi identities of the algebra being satisfied. The deformations of the algebra can be studied using cohomological methods \cite{levynahas}, { see also for example \cite{Gibbons:2007iu}}  . A non-trivial deformation is obtained if it is possible to add covariantly closed but not covariantly exact two forms to the right-hand-sides of the  MC equations \bref{MCBS}. Covariant  exterior differentiation is defined here in terms of the connection 1-forms ${\omega^A}_B={C^A}_{BC}L^C$, where ${C^A}_{BC}$ are the structure constants and $L^C$ the MC 1-forms of the {\it undeformed} algebra. The \Jis imply that the connection is flat: $d{\omega^A}_B+{\omega^A}_C\wedge{\omega^C}_B=0$.

A systematic examination\footnote{
Some of the calculations with forms were done using the Mathematica code
for differential forms developed by S. Bonanos. See: "Graded Exterior Differential Calculus"\cite{bonanos1}. } yields two possible non-trivial deformations
in 4 dimensions, up to redefinitions using covariantly exact one forms.
First one is the $k$-deformation, a supersymmetric extension of the $k$-deformation
of the bosonic Maxwell algebra in \cite{Gomis:2009vm},
\bea
 d {L}_P^\mu+ {L}_M^{\mu\nu} {L}_{P\nu}-{i}
{\overline { {\bL}}}\G^{\mu} {\bL}
&=&  k\,\left( L_Z^{\mu\nu}L_{P\nu} +\frac{i}4 {\overline{{\bL_\Sig}}}\G^{\mu}
{\bL_\Sig}\right),
\nn\\  d {L}_M^{\mu\nu}+  {L}_M^{\mu\rho}\h_{\rho\s} {L}_M^{\s\nu}&=&0,
\nn\\
 d{L}_Z^{\mu\nu}
+ {L}_M^{\mu\rho}\h_{\rho\s}{L}_Z^{\s\nu}+{L}_Z^{\mu\rho}\h_{\rho\s}{L}_M^{\s\nu}- {L}_P^\mu\, {L}_P^\nu- {i}{\overline {  {\bL}}}\G^{\mu\nu}
  {\bL}_\Sig
&=&k\,L_Z^{\mu\rho}{L_{Z\rho}}^\nu,
\nn\\
 d {\bL}^\A+\,\frac14{L}_M^{\mu\nu}(\G_{\mu\nu} {\bL})^\A
+\,  {L}^5\,(\G_5  {\bL})^\A\,
&=&\frac{k}4\,\left( L_Z^{\mu\nu}(\G_{\mu\nu} {\bL})^\A- L_P^{\mu}(\G_{\mu} {\bL_\Sig})^\A\right) ,
\nn\\
d {\bL}_\Sig^{\A}+\frac14 {L}_M^{\mu\nu}(\G_{\mu\nu} {\bL}_\Sig)^\A
+\,  {L}_P^\mu(\G_{\mu}{\bL})^{\A}-\, {L}^5(\G_5  {\bL}_\Sig)^\A
&=&\frac{k}4 L_Z^{\mu\nu}\,(\G_{\mu\nu} {\bL_\Sig})^\A,
\nn\\
d {L}_B- {i}{\overline { {\bL}}}\,\G_{5}\, {\bL}_\Sig&=&0,
\nn\\
 d {L}^5&=&-{k}\,{i}{\overline { {\bL}}}\,\G_{5}\, {\bL}_\Sig,
\label{MCBSsuper}\eea
where $k$ is the deformation parameter having the mass dimension $[k]={2}$.
Note that in contrast to \bref{MCBS}
we cannot have the closed algebra without $L^5$ (chiral symmetry) in the deformed algebra
\bref{MCBSsuper}.
In other words we cannot contract out $L^5$ by  $L^5 \to a\, L^5$ and $a\to 0$
in the last equation of \bref{MCBSsuper}.
The closed MC equation \bref{MCBSsuper} is written in the form of a superalgebra,
\bea
\left[P_\mu,P_{\nu}\right]&=&i\,Z_{\mu\nu},\qquad
\left[P_\mu,Z_{\rho\s}\right]=i\,k\,\h_{\mu[\rho}P_{\s]},
\nn\\
\left[Z_{\mu\nu},Z_{\rho\s}\right]&=&i\,k\,(\h_{\nu[\rho}Z_{|\mu |\s]}-
\h_{\mu[\rho}Z_{|\nu |\s]}),
\nn\\
\left[P_\mu,M_{\rho\s}\right]&=&-i\,\h_{\mu[\rho}P_{\s]},\qquad
\left[Z_{\mu\nu},M_{\rho\s}\right]=-i\,\h_{\nu[\rho}Z_{|\mu |\s]}+i\,
\h_{\mu[\rho}Z_{|\nu |\s]},
\nn\\
\left[M_{\mu\nu},M_{\rho\s}\right]&=&-i\,\h_{\nu[\rho}M_{| \mu |\s]}+i\,
\h_{\mu[\rho}M_{|\nu |\s]},
\\
\{{\bQ}_\A,{\bQ}_\B\}&=&2\,(C\G^{\mu})_{\A\B}P_\mu,\qquad
\{{\bS}_\A,{\bS}_\B\}=\frac12\,k\,(C\G^{\mu})_{\A\B}P_\mu,
\nn\\
\{{\bQ}_\A,{\bf\Sigma}_{\B}\}&=&\frac12(C\G^{\mu\nu})_{\A\B}\,Z_{\mu\nu}\,+\, (C\G_5)_{\A\B}\,({\rB}-k\,B_5),
\\
\left[P_\mu,{\bQ}_\A\right]&=&-i\,\bS_{\B}(\G_{\mu}{)^{\B}}_\A,\qquad
\left[P_\mu,{\bS}_\A\right]=-\frac{i}4\,k\,\bQ_{\B}(\G_{\mu}{)^{\B}}_\A,
\nn\\
\left[Z_{\mu\nu},{\bQ}_\A\right]&=&\frac{i}2k\,(\bQ\G_{\mu\nu})_\A,\qquad
\left[Z_{\mu\nu},{\bS}_\A\right]=\frac{i}2k\,(\bS\G_{\mu\nu})_\A,
\nn\\
\left[B_5,{\bQ}_\A\right]&=&-i\,(\bQ\G_{5})_\A,\qquad
\left[B_5,{\bS}_\A\right]=\,i\,(\bS\G_{5})_\A,
\nn\\
\left[M_{\rho\s},\bQ_{\A}\right]&=&-\frac{i}2(\bQ \G_{\rho\s})_{\A},\qquad
\left[M_{\rho\s},\bS_{\A}\right]=-\frac{i}2(\bS \G_{\rho\s})_{\A}.
\label{kalgebra}
\eea
\vs

The second $s$-deformation is
\bea
 d {L}_P^\mu+ {L}_M^{\mu\nu} {L}_{P\nu}-{i}{\overline { {\bL}}}\G^{\mu} {\bL}
&=& {s}\,L^5\,L_P^\mu,
\nn\\
 d {L}_M^{\mu\nu}+  {L}_M^{\mu\rho}\h_{\rho\s} {L}_M^{\s\nu}&=&
0,
\nn\\
 d{L}_Z^{\mu\nu}
+ {L}_M^{\mu\rho}\h_{\rho\s}{L}_Z^{\s\nu}+{L}_Z^{\mu\rho}\h_{\rho\s}{L}_M^{\s\nu}- {L}_P^\mu\, {L}_P^\nu- {i}{\overline {  {\bL}}}\G^{\mu\nu}
  {\bL}_\Sig
&=&{2s}\,L^5\,L_Z^{\mu\nu},
\nn\\
 d {\bL}^\A+\,\frac14{L}_M^{\mu\nu}(\G_{\mu\nu} {\bL})^\A
+\,  {L}^5\,(\G_5  {\bL})^\A\,
&=&\frac{1}2s\,L^5\,{\bL}^{\A},
\nn\\
d {\bL}_\Sig^{\A}+\frac14 {L}_M^{\mu\nu}(\G_{\mu\nu} {\bL}_\Sig)^\A
+\,  {L}_P^\mu(\G_{\mu}{\bL})^{\A}-\, {L}^5(\G_5  {\bL}_\Sig)^\A
&=&\frac{3}2s\,L^5\,{\bL}_\Sig^{\A},
\nn\\
d {L}_B- {i}{\overline { {\bL}}}\,\G_{5}\, {\bL}_\Sig&=&2\,s\,L^5\,{L}_B,
\nn\\
 d {L}^5&=&0.
\label{MCBSb}\eea
where $s$ is the dimensionless deformation parameter.

%%%%%%%%%%%%%%%%%%%%%%%%%%%
\subsection{$k$-deformation }
We describe the $k$-deformation of the \sM \  in terms of known super algebras.
We write
\be
L_M-k\,L_Z= L_{\CM},\qquad
L_M=L_{\CJ},\qquad L_B+\frac1kL^5=\frac1k L_\CB,\qquad L^5=L_{\CB^5},
\label{decouple}\ee
and rescale using $k=\pm\frac{1}{R^2}, (R>0),$
\be
L_P^\mu=R\,L_\CP^\mu,\qquad
\bL^\A=\sqrt{R}\,\7\bL^\A,\qquad
\bL^\A_\Sig=\sqrt{R^3}\,\7\bL^\A_\Sig
\ee
so that all one forms are dimensionless.
 Correspondingly the relations of the new generators to those of
$k$-deformed \sM\ are found by comparing \bref{newalgebraMC} with
\bea
\W&=&
\CP_\mu L_\CP^\mu+\frac12 \CM_{\mu\nu} L_\CM^{\mu\nu}+\frac12 \CJ_{\mu\nu} L_\CJ^{\mu\nu}+\CB L_\CB +\CB_5 L_{\CB^5} +\7\bQ_\A\7\bL^\A+\7\bS_\A\7\bL_\Sig^\A,
\eea
where
\bea
M_{\mu\nu}&=&\CM_{\mu\nu}+\CJ_{\mu\nu},\quad
Z_{\mu\nu}=\mp\,\frac{1}{R^2}\, \CM_{\mu\nu},\quad B =\pm\frac1{R^2}\CB ,\quad
B_5=\CB_5 +\CB,\nn\\
P_{\mu}&=&\frac1R\CP_{\mu},\qquad
\bQ_{\A}=\frac{1}{\sqrt{R}}\,\7\bQ_{\A} ,\qquad
\bS_{\A}=\frac{1}{\sqrt{R^3}}\,\7\bS_{\A} .
\label{relgenSM3}\eea

The MC equations {\bref{MCBSsuper}} decompose to that of $\SO(3,1)$ of $\CJ$
\bea
 d {L}_\CJ^{\mu\nu}+  {L}_\CJ^{\mu\rho}\h_{\rho\s} {L}_\CJ^{\s\nu}&=&0,
\eea
the central charge $\CB$,
\be
d {L}_\CB=0
\ee
and a superalgebra of $(\CP,\CM,\7\bQ,\7\bS,\CB_5)$,
\bea
 d {L_\CP}^\mu+ {L}_\CM^{\mu\nu} {L_\CP}_\nu - {i}
{\overline { {\7\bL}}}\G^{\mu} {\7\bL}
&=&\pm\,\frac{i}4 {\overline{{\7\bL_\Sig}}}\G^{\mu}{\7\bL_\Sig},
\nn\\
 d{L}_\CM^{\mu\nu}+ {L}_\CM^{\mu\rho}\h_{\rho\s}{L}_\CM^{\s\nu}
&=&\mp {L}^\mu_\CP\, {L}^\nu_\CP\mp\, {i}\,{\overline {  {\7\bL}}}\G^{\mu\nu}
{\7\bL}_\Sig,
\nn\\
 d {\7\bL}^\A+\,\frac14{L}_\CM^{\mu\nu}(\G_{\mu\nu} {\7\bL})^\A
+\,  L_{\CB^5}\,(\G_5  {\7\bL})^\A\,
&=&\mp\, L_\CP^{\mu}\frac{1}4\,(\G_{\mu} {\7\bL_\Sig})^\A,
\nn\\
d {\7\bL}_\Sig^{\A}+\frac14 {L}_\CM^{\mu\nu}(\G_{\mu\nu} {\7\bL}_\Sig)^\A
+\,  {L_\CP^\mu}(\G_{\mu}{\7\bL})^{\A}-\,L_{\CB^5}(\G_5  {\7\bL}_\Sig)^\A
&=&0,
\nn\\
 d {L_{\CB^5}}&=&\mp\,{i}{\overline { {\7\bL}}}\,\G_{5}\, {\7\bL}_\Sig,
\label{MCBSsuper2}\eea
where upper signs correspond to $k>0$ and lower ones to $k<0$.
Both cases ($k>0$ and $k<0$) will be discussed in detail below.

%%%%%%%%%%%%%%%%%%%%%%%%%%%%%%%%%%%%%%%%%%%

\subsubsection{Anti-de Sitter case: $k>0$ }

We shall show that
the superalgebra of $(\CP,\CM,\7\bQ,\7\bS,\CB_5)$ in \bref{MCBSsuper2}
for $k>0$ is isomorphic to $OSp(2|4)$, which is N=2, D=4 anti-de-Sitter
superalgebra with bosonic subalgebras $\SO(2)$ and $Sp(4)\sim \SO(3,2)$. The MC equation  \bref{MCBSsuper2}
for $k>0$ is written in $\SO(3,2)$ covariant form as
\bea
 d{L}_\CM^{{\8\mu}{\8\nu}}+ {L}_\CM^{{\8\mu}\8\rho}\h_{\8\rho\8\s}{L}_\CM^{\8\s{\8\nu}}
+\frac{i}2\,{\overline {{\bL^i}}}\Gam^{{\8\mu}{\8\nu}} {\bL}^i
&=&0,
\nn\\
 d {\bL}^{\A i}+\,\frac14{L}_\CM^{{\8\mu}{\8\nu}}(\Gam_{{\8\mu}{\8\nu}} {\bL})^{\A i}
+\,{L}_{\CB^5}\,\ep^{ij}{\bL}^{\A j}\,
&=&0 ,
\nn\\
 d {L}_{\CB^5}-\,\frac{i}2 \,{\overline { {\bL}}}^i\,\ep^{ij}\,{\bL}^j&=&0,
\label{MCBSsuperAdS5}\eea
where ${\8\mu},{\8\nu}=0,1,2,3,4$ are $\SO(3,2)$ indices with the flat metric $\h_{{\8\mu}{\8\nu}}=(-,+++,-)$ and  $\ep^{12}=-\ep^{21}=1$.
$L_\CM^{\mu\nu}$'s are  $\SO(3,1)$ components of  $L_\CM^{{\8\mu}{\8\nu}}$ and
\be
L_\CP^{\mu}= L_\CM^{\mu 4}.
\ee
We have also introduced $\SO(3,2)$ Majorana spinors
\be
\bL^{i\A}=\pmatrix{\bL^{1\A}\cr\bL^{2\A}}=
\pmatrix{\7\bL^{\A}+\frac12\7\bL_\Sig^\A\cr -\Gam_4(\7\bL^{\A}-\frac12\7\bL_\Sig^\A)}
\ee
and the $\SO(3,2)$ gamma matrices $\Gam_{\8\mu}$ given in appendix A.
Recall that all gamma matrices are taken to be real in the Majorana representation.
\bref{MCBSsuperAdS5} shows that
$\bL^{i\A},(i=1,2)$ is a $\SO(2)$ doublet under  rotations generated by $\CB^5$.
The dual superalgebra
of the MC equation \bref{MCBSsuperAdS5} is OSp(2$|$4):
\bea
\left[\CM_{{\8\mu}{\8\nu}},\CM_{\8\rho\8\s}\right]&=&-i\,\h_{{\8\nu}[\8\rho}\CM_{|{\8\mu}|\8\s]}+i\,
\h_{{\8\mu}[\8\rho}\CM_{|{\8\nu}|\8\s]},
\nn\\
\{{\CQ}_{\A i},{\CQ}_{\B j}\}&=&-\frac{1}{2}\,\D_{ij}\,(C\Gam^{{\8\mu}{\8\nu}})_{\A\B}\CM_{{\8\mu}{\8\nu}}+\ep_{ij}\,(C)_{\A\B}\CB_5,
\nn\\
\left[\CM_{{\8\mu}{\8\nu}},\CQ_{\A i}\right]&=&-\frac{i}2(\CQ_i \Gam_{{\8\mu}{\8\nu}})_{\A},
\nn\\
\left[\CB_5,{\CQ}_{\A i}\right]&=&i\,\ep_{ij}\,\CQ_{\A j}.
\label{superAdS5}\eea
Thus the $k$-deformed \sM \, for $k>0$ is isomorphic to
$OSp(2|4)\oplus \SO(3,1)\oplus U(1)$ with the generators related by \bref{relgenSM3} ,
% ( checked in nb0117DefMaxPB )
\bea
\CM_{\mu 4}&=&R\,P_\mu,\qquad
\CM_{\mu\nu}=-R^2\, Z_{\mu\nu},\qquad
\CB_5=B_5  -R^2\,B,\nn\\
\CQ_{\A1}&=& \frac{\sqrt{R}}{2}\bQ_{\A}+\sqrt{R^3}\,\bS_{\A},\qquad
\CQ_{\A2}= (\frac{\sqrt{R}} {2}\bQ_{\B}-\sqrt{R^3}\,\bS_{\B}){\Gam_4^\B}_\A,
\nn\\
\CJ_{\mu\nu}&=&M_{\mu\nu}+R^2\, Z_{\mu\nu},\qquad
\CB =R^2\,B .
\label{relgenSM3Rplus}
\eea
The unconventional feature of the contraction \bref{relgenSM3Rplus} is exhibited in the formula for $\CJ_{\mu\nu}$ in which the direct sum structure of two Lorentz subalgebras belonging to $OSp(2|4)\oplus \SO(3,1)$ is not respected. Such nonstandard contractions generating nontrivial cohomologies and semidirect sum of algebras were considered in D=3 (see \cite{deAzcarraga:2002xi} sect.8, \cite{Cangemi:1992ri}).
\vs

%%%%%%%%%%%%%%%%%%%%%%%%%%%%%%%%%%
\subsubsection{de Sitter case: $k<0$}

The superalgebra of $(\CP,\CM,\7\bQ,\7\bS,\CB_5)$ in \bref{MCBSsuper2}
for $k=-\frac{1}{R^2}<0$ case  is,  as we will see,  the N=1 super dS algebra \usp.
In order to consider this case we introduce $\SO(4,1)$ metric and
gamma matrices $\Gam_{\8\mu}$ as (see appendix A)
\be
\Gam_\mu=\G_\mu\G_5,\quad
\Gam_4=i\,\G_5,\qquad \h_{{\8\mu}{\8\nu}}=(-,+++;+),\quad {\8\mu},{\8\nu}=0,1,2,3,4.
\ee
In contrast to the AdS case these gamma matrices are not real but satisfy
\be
\Gam_{\8\mu}^*=-\Gam_4\,\Gam_{\8\mu}\,\Gam_4,\qquad
\Gam_{{\8\mu}{\8\nu}}^*=\Gam_4\,\Gam_{{\8\mu}{\8\nu}}\,\Gam_4.
\label{stargamma}\ee
Using them we can rewrite \bref{MCBSsuper2} in $\SO(4,1)$ covariant form,
\bea
 d{L}_\CM^{{\8\mu}{\8\nu}}+ {L}_\CM^{{\8\mu}\8\rho}\h_{\8\rho\8\s}{L}_\CM^{\8\s{\8\nu}}
+\frac12\,{\overline {{\bL^i}}}\Gam^{{\8\mu}{\8\nu}} {\bL}^i
&=&0,
\nn\\
 d {\bL}^{\A i}+\,\frac14{L}_\CM^{{\8\mu}{\8\nu}}(\Gam_{{\8\mu}{\8\nu}} {\bL})^{\A i}
+\,{L_{\CB^5}}\,\ep^{ij}{\bL}^{\A j}\,
&=&0 ,
\nn\\
 d {L_{\CB^5}}-\,\frac12 \,{\overline { {\bL}}}^i\,\ep^{ij}\,{\bL}^j&=&0.
\label{MCBSsuperdS5}\eea
Here
\be
L_\CM^{\mu 4}=L_\CP^\mu,\qquad \bL^{i\A}=\pmatrix{\bL^{1\A}\cr\bL^{2\A}}=
\pmatrix{\7\bL^{\A}-\frac{i}2\7\bL_\Sig^\A\cr -i\,\Gam_4(\7\bL^{\A}+\frac{i}2\7\bL_\Sig^\A)}.
\label{smdoublet}\ee
The $\bL^{i\A}$'s are not Majorana spinors but are symplectic Majorana spinors
satisfying the condition,  see for example  \cite{VanProeyen:1999ni},
\bea
{\bL^{\A i}}^\dagger&=&i\ep^{ij} \,{{(\Gam_4)}^\A}_\B\,\bL^{\B j}.
\label{SML}\eea
The  symplectic Majorana spinors \bref{smdoublet} transform as an $\SO(2)$  doublet under rotation by $\CB^5$.
\vs

The dual superalgebra of the MC equation \bref{MCBSsuperdS5} is
\bea
\left[\CM_{{\8\mu}{\8\nu}},\CM_{\8\rho\8\s}\right]&=&-i\,\h_{{\8\nu}[\8\rho}\CM_{|{\8\mu}|\8\s]}+i\,
\h_{{\8\mu}[\8\rho}\CM_{|{\8\nu}|\8\s]},
\nn\\
\{{\CQ}_{\A i},{\CQ}_{\B j}\}&=&\frac{i}{2}\,\D_{ij}\,(C\Gam^{{\8\mu}{\8\nu}})_{\A\B}\CM_{{\8\mu}{\8\nu}}
-i\ep_{ij}\,(C)_{\A\B}\CB_5,
\nn\\
\left[\CM_{{\8\mu}{\8\nu}},\CQ_{\A i}\right]&=&-\frac{i}2(\CQ_i \Gam_{{\8\mu}{\8\nu}})_{\A},
\nn\\
\left[\CB_5,{\CQ}_{\A i}\right]&=&i\,\ep_{ij}\,\CQ_{\A j}.
\label{superdS5}\eea
Here $\CM_{{\8\mu}{\8\nu}}$ and $\CB_5$ are real (Hermitian) while ${\CQ}_{\A i}$ are complex and satisfy the symplectic SU(2) Majorana or quaternionic condition
\be
\CQ_{\A i}^\dagger=-\CQ_{\B j}\,i\ep^{ij} \,{{(\Gam_4)}^\B}_\A.
\label{symMajo2}\ee
The superalgebra \bref{superdS5}  is N=1, D=4 de-Sitter superalgebra \usp \ discussed in \cite{Lukierski:1984it}\cite{Pilch:1984aw} and Appendix B. We can conclude that  the  $k$-deformed \sM \  for $k<0$ is isomorphic to
\usp $\oplus \SO(3,1)\oplus U(1)$ with the generators related by  
% ( checked in nb0117DefMaxPB )
\bea
\CM_{\mu 4}&=&R\,P_\mu,\qquad
\CM_{\mu\nu}=R^2\, Z_{\mu\nu},\qquad
\CB_5=B_5  +R^2\,B,\nn\\
\CQ_{\A1}&=& \frac{\sqrt{R}}{2}\,\bQ_{\A}+i\,\sqrt{R^3}\,\bS_{\A},\qquad
\CQ_{\A2}= (\frac{i\sqrt{R}} {2}\,\bQ_{\B}+\sqrt{R^3}\,\bS_{\B}){\Gam_4^\B}_\A,
\nn\\
\CJ_{\mu\nu}&=&M_{\mu\nu}-R^2\, Z_{\mu\nu},\qquad
\CB =-R^2\,B .
\label{relgenSM3Rmin}
\eea

We add that the contractions of the deformed algebras
$OSp(2|4)\oplus \SO(3,1)\oplus U(1)$ and \usp $\oplus \SO(3,1)\oplus U(1)$
%e.g. in the contraction procedure one can eliminate
are not unique but other contractions  are possible to produce results different from the Maxwell superalgebra \bref{newalgebra}.
For example, one could  get N=2 Poincar{\`e} superalgebra$\oplus \SO(3,1)\oplus U(1)^2$.

%%%%%%%%%%%%%%%%%%%%%%%%%%%%%%%%%
\subsection{$s$-deformation}

 From  the MC equation \bref{MCBSb} it follows  that
only the action of the $B_5$ generator is changed in comparison with the undeformed algebra \bref{newalgebra}. Denoting the $s$-deformed generator by $B_5^{(s)}$, one gets
\bea
\left[B_5^{(s)}, P_{\mu}\right]&=&i\,s\,P_{\mu},\quad\left[B_5^{(s)}, Z_{\mu\nu}\right]=i \,2\,s\,Z_{\mu\nu},\quad
\left[B_5^{(s)}, B\right]=i \,2\,s\, B,\quad \left[B_5^{(s)},M_{\rho\s}\right]=0,
\nn\\
\left[B_5^{(s)},{\bQ}_\A\right]&=&i \,(\bQ(\frac s2-\gamma_5))_\A,\qquad
\left[B_5^{(s)},{\bS}_\A\right]=\,i\,(\bS(\frac {3\,s}2+\gamma_5))_\A.
\label{dilatation2}\eea

 The generator $B_5^{(s)}$ acts on all generators as a dilatation operator and produces as well the  chiral rotations for the fermionic generators. Note also that the generator $B$, after the $s$-deformation, ceases to be central. From \bref{newalgebra}, \bref{Daction} and \bref{dilatation2} it follows that the deformed chiral generator $B_5^{(s)}$ can be represented as
\be
B_5^{(s)}=B_5+s\,\CD.\label{B5sDef}
\ee
Because of $[\CD,\,B_5]=0$, the relation \bref{B5sDef} selects  a one-parameter subalgebra
from the two-dimensional Abelian subalgebra $(\CD,\,B_5)$ of the Maxwell-Weyl superalgebra. We conclude therefore that the $s$-deformed D=4 \sM \ is embedded in the Maxwell-Weyl superalgebra considered in section 2.

%%%%%%%%%%%%%%%%%%%%%%%%%%%%%%%%%%%%%%%%%%%%%%%%%%%%%%%%%%%%%%55
\section{D=3 \sM \, and its deformations}

\subsection{D=3 \sM }

 The D=3 Majorana spinors are two-component, and $\SO(2,1)$ gamma matrices in Majorana representation can be expressed in terms of the Pauli matrices $\s_i$ as follows
\be\label{sigmamatrices}
C=\G_0=i\s_2,\quad  \G_1=\s_1,\quad  \G_2=\s_3.
\ee
Because the chiral product of the three matrices \bref{sigmamatrices} is the identity matrix
we cannot accommodate the generators $B$ and $B_5$ in D=3 Maxwell superalgebra, which by analogy with the formulae \bref{newalgebra} takes the following form
\bea
\left[P_\mu,P_{\nu}\right]&=&i\,Z_{\mu\nu},\qquad\qquad
\left[P_\mu,{\bQ}_\A\right]=-i\,\bS_{\B}(\G_{\mu}{)^{\B}}_\A,
\nn\\
\{{\bQ}_\A,{\bQ}_\B\}&=&2\,(C\G^{\mu})_{\A\B}P_\mu,\qquad
\{{\bQ}_\A,{\bf\Sigma}_{\B}\}=\frac12(C\G^{\mu\nu})_{\A\B}\,Z_{\mu\nu},
\nn\\
\left[P_\mu,M_{\rho\s}\right]&=&-i\,\h_{\mu[\rho}P_{\s]},
\qquad
\left[Z_{\mu\nu},M_{\rho\s}\right]=-i\,\h_{\nu[\rho}Z_{|\mu |\s]}+i\,
\h_{\mu[\rho}Z_{|\nu |\s]},
\nn\\
\left[M_{\rho\s},\bQ_{\A}\right]&=&-\frac{i}2(\bQ \G_{\rho\s})_{\A},\qquad
\left[M_{\rho\s},\bS_{\A}\right]=-\frac{i}2(\bS \G_{\rho\s})_{\A},
\nn\\
\left[M_{\mu\nu},M_{\rho\s}\right]&=&-i\,\h_{\nu[\rho}M_{| \mu |\s]}+i\,
\h_{\mu[\rho}M_{|\nu |\s]}.
\label{newalgebra3D}\eea
The Jacobi identities of this algebra hold using the identity
$(C\G_\mu)_{(\A\B}(C\G^\mu)_{\gam\D)}=0$, for totally symmetric $(\A\B\gam\D)$.
 We would like to mention that the D=3 Maxwell algebra was considered earlier under the name of extended Poincar{\`e} algebra \cite{Cangemi:1992ri}\cite{Duval:2008tr}. The D=3 \sM \ \bref {newalgebra3D} is given here for the first time.

\vs

\subsection{Deformations of D=3 \sM }

In D=3 dimensions  there are no $B$ and $B_5$ generators and there is
no $s$-deformation. On the other hand the presence of the totally antisymmetric
tensor $\epsilon^{\mu\nu\rho}$ 
leads to a new deformation, as in the bosonic case ($b$-deformation
in \cite{Gomis:2009vm}), in addition to the $k$-deformation \bref{MCBSsuper}:
\bea
d {L}_M^{\mu\nu}+  {L}_M^{\mu\rho}\h_{\rho\s} {L}_M^{\s\nu}&=&0,
\nn\\
 d {L}_P^\mu+ {L}_M^{\mu\nu} {L}_{P\nu}-{i}
{\overline { {\bL}}}\G^{\mu} {\bL}
&=&  c\,{\ep^\mu}_{\rho\s}L_P^{\rho}L_P^\s,+k
\left( L_Z^{\mu\nu}L_{P\nu} +\frac{i}4 {\overline{{\bL_\Sig}}}\G^{\mu}
{\bL_\Sig}\right),
\nn\\
 d{L}_Z^{\mu\nu} + {L}_M^{\mu\rho}\h_{\rho\s}{L}_Z^{\s\nu}+{L}_Z^{\mu\rho}\h_{\rho\s}{L}_M^{\s\nu}- {L}_P^\mu\, {L}_P^\nu&-&{i}{\overline {  {\bL}}}\G^{\mu\nu}
  {\bL}_\Sig =-\frac{c}2\,i\,{\overline{{\bL_\Sig}}}\G^{\mu\nu}
{\bL_\Sig} +k\,L_Z^{\mu\rho}{L_{Z\rho}}^\nu,
\nn\\
 d {\bL}^\A+\,\frac14{L}_M^{\mu\nu}(\G_{\mu\nu} {\bL})^\A
&=&-c\,L_P^\mu(\G_{\mu}{\bL})^\A+
\frac{k}4\,\left( L_Z^{\mu\nu}(\G_{\mu\nu} {\bL})^\A-L_P^{\mu}(\G_{\mu}
{\bL_\Sig})^\A \right) ,
\nn\\
d {\bL}_\Sig^{\A}+\frac14 {L}_M^{\mu\nu}(\G_{\mu\nu} {\bL}_\Sig)^\A
+ {L_P}^\mu(\G_{\mu}{\bL})^{\A}\,
&=&\frac{k}4\,L_Z^{\mu\nu}(\G_{\mu\nu} {\bL_\Sig})^\A ,
\label{MCBSsuper0}\eea
where $c,\,k$ are the deformation parameters having the mass dimension
$ [k]=2,\; [c]={1}.$
$\bL$ and $\bL_\Sig$ are 2-component Majorana spinors.
We will examine the symmetry structure 
of these deformations in the three cases,
(1) $k^+$-deformation for $k>0$, (2) $k^-$-deformation for $k<0$
and (3) $c$-deformation.
\vs
\subsubsection{ $k^+$-deformation}
 In the case $k>0,\; c=0$ the MC equation is written as
\bea
dL_\CJ^\mu+\frac12{\ep^\mu}_{\nu\rho}L_\CJ^\nu L_\CJ^\rho&=&0,
\nn\\
dL_\pm^{\mu}+\frac12{\ep^\mu}_{\nu\rho}L_\pm^\nu L_\pm^\rho\mp
 {i}{{\bL}_\pm}C\G^{\mu}{\bL}_\pm &=&0,\qquad
 d{\bL}_\pm^{\A}-\,\frac12L_\pm^{\mu}(\G_{\mu} {\bL}_\pm)^\A=0,
\eea
where
\be
L_\CJ^\mu=\frac12{\ep^\mu}_{\nu\rho}L_{M}^{\nu\rho},\quad
L_\pm^{\mu}=\frac12{\ep^\mu}_{\nu\rho}(L_{M}^{\nu\rho}-\frac{L_{Z}^{\nu\rho}}
{R^2})\pm\frac{L_{P}^{\mu}}{R} ,\quad
{\bL}_\pm^{\A}=\frac{{\bL}^{\A}}{R^{1/2}}\mp \frac{\bL_{\Sig}^{\A}}{2\,R^{3/2}}
\label{relkplus}\ee
and $\bL_\pm$ are two independent Majorana spinors.

The dual algebra for the generators, defined by using the MC one form
\bea
\W&=&\CJ_{\mu}L_\CJ^{\mu}+ \CM_{\mu}^+L_+^{\mu}+\CM_{\mu}^-L_-^{\mu}+
{\CQ_\A^+}\bL_+^{\A}+{\CQ_\A^-}\bL_-^{\A}
\eea
becomes
\bea
\left[\CJ_\mu,\CJ_{\nu}\right]&=&-i\ep_{\mu\nu\rho}\CJ^{\rho},
\nn\\
\left[\CM_\mu^\pm,\CM_{\nu}^\pm\right]&=&-i\ep_{\mu\nu\rho}\CM^{\pm\rho},
\qquad
\left[\CM_{\mu}^\pm,\CQ_{\A}^\pm\right]=\frac{i}2(\CQ^\pm \G_{\mu})_{\A},
\nn\\
\{{\CQ}_\A^\pm,{\CQ}_\B^\pm\}&=&\pm\,2\,(C\G^{\mu})_{\A\B}\CM_\mu^\pm.
\label{kminusalgebra3D}
\eea
 Here, due to the relation $\sum_\A \CQ_\A^\pm \CQ_\A^\pm=-\sum_\A|\CQ_\A^\pm|^2=\pm 2 \CM_0^\pm$,
we obtain the constraints $\CM_0^+\leq 0, \CM_0^-\geq 0$.
The $\CJ_\mu$ is the $\SO(2,1)$ generator and  $(\CM_\mu^\pm,{\CQ}_\A^\pm)$
generates a pair of real superalgebras $OSp^{\pm}(1|2)$.
Then the $k^+$-deformed super algebra is a direct sum:

\be
k^+{\rm -deformed\; superalgebra} = \SO(2,1)\oplus OSp^+(1|2)\oplus OSp^-(1|2).
\ee
\vs

\subsubsection{ $k^-$-deformation }
In the case $k<0, \; c=0$ the MC equation is written as
\bea
dL_\CJ^\mu+\frac12{\ep^\mu}_{\nu\rho}L_\CJ^\nu L_\CJ^\rho&=&0,
\nn\\
dL_\pm^{\mu}+\frac12{\ep^\mu}_{\nu\rho}L_\pm^\nu L_\pm^\rho\pm
{{\bL}_\pm}C\G^{\mu}{\bL}_\pm &=&0,\quad
d{\bL}_\pm^{\A}-\,\frac12\,L_\pm^{\mu}(\G_{\mu} {\bL}_\pm)^\A=0,
\eea
where
\be
L_\CJ^\mu=\frac12{\ep^\mu}_{\nu\rho}L_{M}^{\nu\rho},\quad
L_\pm^{\mu}=\frac12{\ep^\mu}_{\nu\rho}(L_{M}^{\nu\rho}+\frac{L_{Z}^{\nu\rho}}
{R^2})\pm\,i\,\frac{L_{P}^{\mu}}{R} ,\quad
{\bL}_\pm^{\A}=\frac{\bL^{\A}}
{R^{1/2}}\mp \,i\,\frac{\bL_{\Sig}^{\A}}
{2\,R^{3/2}},
\ee and $L_+^{\mu}=L_-^{\mu\dagger},\,{\bL}_+^{\A}={\bL}_-^{\A\dagger}.$
The dual algebra for the generators, defined by using the MC one form
\bea
\W&=&\CJ_{\mu}L_\CJ^{\mu}+ \CM_{\mu}^+L_+^{\mu}+\CM_{\mu}^-L_-^{\mu}+
{\CQ_\A^+}\bL_+^{\A}+{\CQ_\A^-}\bL_-^{\A}
\label{Wkminus}\eea
becomes
\bea
\left[\CJ_\mu,\CJ_{\nu}\right]&=&-i\ep_{\mu\nu\rho}\CJ^{\rho},
\nn\\
\left[\CM_\mu^\pm,\CM_{\nu}^\pm\right]&=&-i\ep_{\mu\nu\rho}\CM^{\pm\rho},
\qquad
\left[\CM_{\mu}^\pm,\CQ_{\A}^\pm\right]=\frac{i}2(\CQ^\pm \G_{\mu})_{\A},
\nn\\
\{{\CQ}_\A^\pm,{\CQ}_\B^\pm\}&=&\pm\,2\,i\,(C\G^{\mu})_{\A\B}\CM_\mu^\pm.
\label{kplusalgebra3D}
\eea
Note there is a factor $i$ in the $\{\CQ,\CQ\}$ anticommutator because of
the hermiticity requirement
\be
(\CM_\mu^-)^\dagger=\CM_\mu^+,\qquad
(\CQ_\A^-)^\dagger=-\CQ_\A^+.
\ee
Because $\CQ^+_\A$ and $\CQ^-_\B$ are anticommuting
\be
\{\CQ^+_\A,\CQ^-_\B\}=-\{\CQ^+_\A,{\CQ^+_\B}^\dagger\}=0,
\ee
one obtains that $\sum_\A|\CQ_\A^\pm|^2=0$ and representations at quantum level  of the superalgebra \bref{kplusalgebra3D} lead necessarily to indefinite metric and ghost states 
\cite{Pilch:1984aw}\cite{Lukierski:1984it}.
 The algebra is a direct sum of \SO(2,1) with generator $\CJ$,
$OSp(1|2;C)$  with complex generators $\CM_+,\CQ_+$ and
${\overline {OSp(1|2;C)}}$ with their conjugate generators
$\CM_-,\CQ_-$. 
We obtain
\be
k^-{\rm -deformed\; superalgebra} = \SO(2,1)\oplus OSp(1|2;C)\oplus {\overline {OSp(1|2;C)}}.
\label{kminusdef}\ee
\vs

\subsubsection{ $c$-deformation }
The $c$-deformation is the case $k=0$ given by
\bea
dL_\CJ^\mu+\frac12{\ep^\mu}_{\nu\rho}L_\CJ^\nu L_\CJ^\rho&=&-2c\,{i}{\overline { {\7\bL}}}\G^{\mu} {
\7\bL},\qquad d {\7\bL}^\A-\,\frac12{L}_\CJ^{\mu}(\G_{\mu} {\7\bL})^\A=0,
\nn\\
dL_\CM^\mu+\frac12{\ep^\mu}_{\nu\rho}L_\CM^\nu L_\CM^\rho&=&0,\nn\\
dL_\CP^\mu+{\ep^\mu}_{\nu\rho}L_\CM^\nu L_\CP^\rho&=&-\frac{i}2\,{\overline { {\7\bL_\Sig}}}\G^{\mu} {\7\bL_\Sig},\qquad
d {\7\bL}_\Sig^{\A}-\frac12 {L}_\CM^{\mu}(\G_{\mu} {\7\bL}_\Sig)^\A=0.
\label{scdefnew}
\eea
where
\be
L_\CJ^\mu=-2cL_P^\mu+\frac12{\ep^\mu}_{\nu\rho}L_M^{\nu\rho},\quad
L_\CP^\mu=c\,\frac12{\ep^\mu}_{\nu\rho}L_Z^{\nu\rho}-\frac12L_P^\mu,\quad
L_\CM^\mu=\frac12{\ep^\mu}_{\nu\rho}L_M^{\nu\rho},
\ee
and
\be
\7\bL^\A=\bL^\A,\qquad \7\bL_\Sig^\A=\bL^\A-c\,\bL_\Sig^\A,\qquad
 [\7\bL]=[\7\bL_\Sig]={-1/2}.
\ee
The dual algebra for the generators, defined by using the MC one form
\bea
\W&=&\CJ_{\mu}L_\CJ^{\mu}+{\7\bQ_\A}\7\bL^\A+\CM_{\mu}L_\CM^{\mu}+
\CP_{\mu}L_\CP^{\mu}+{\7\bS_\A}\7\bL_\Sig^{\A}\eea
is
\bea
\left[\CJ_\mu,\CJ_{\nu}\right]&=&-i\ep_{\mu\nu\rho}\CJ^{\rho},\qquad
\{{\7\bQ}_\A,{\7\bQ}_\B\}=\,-4\,c\,(C\G^{\mu})_{\A\B}\CJ_\mu,
\nn\\
\left[\CM_\mu,\CM_{\nu}\right]&=&-i\ep_{\mu\nu\rho}\CM^{\rho},
\qquad
\left[\CM_{\mu},\7\bS_{\A}\right]=\frac{i}2(\7\bS \G_{\mu})_{\A},
\nn\\
\left[\CP_\mu,\CM_{\nu}\right]&=&-i\ep_{\mu\nu\rho}\CP^{\rho},\qquad
\{{\7\bS}_\A,{\7\bS}_\B\}=-\,(C\G^{\mu})_{\A\B}\CP_\mu.
\label{calgebra3D}
\eea
Here ${\7\bQ}_\A$ and ${\7\bS}_\A$ are independent Majorana spinors.
The algebra with generators  $(\CJ,\7Q)$ is $OSp(1|2)$ and one with
$(\CM,\CP,\7\Sigma)$ is N=1, D=3 Poincar{\`e} superalgebra, i.e.,
\be
c-{\rm deformed\; superalgebra}=OSp(1|2)\oplus ({\rm N=1\; D=3 \;superPoincar{e}}).
\ee
The $c$-deformation is the supersymmetrization of the degenerate case of the
two parameter deformation of D=3 bosonic Maxwell algebra \cite{Gomis:2009vm}.

%%%%%%%%%%%%%%%%%%%%%%%%%%%%%%%%%%%%%%%%%%%%%%%%%%%%%%%%%%%%%%%%%%%%%%%
\section{Casimir Operators of the \sMs}
In this section we discuss how the Casimir operators of \sMs \ can be obtained as contractions of their deformed counterparts and their subalgebras.
%\hfill (see r8629SOnCasimir.tex)
\subsection{Casimir Operators of the \sMs \ in D=4}

In four dimensions there are four Casimir operators in the bosonic Maxwell algebra \cite{Schrader:1972zd}\cite{Soroka:2004fj},
\bea C_1&=&\frac12\,Z_{\mu\nu}^2,
\qquad C_2=\frac12(Z_{\mu\nu}\7Z^{\mu\nu}),
\nn\\
C_3&=&P_\mu^2+M_{\mu\nu}Z^{\mu\nu},\qquad 
C_4=(P^{\nu}\7Z_{\nu\mu})^2+ \frac14(Z_{\mu\nu}\7Z^{\mu\nu})\,(M_{\rho\s}\7Z^{\rho\s}),
\label{4Casimir}\eea
where  $\7Z^{\mu\nu}=\frac12\ep^{\mu\nu\rho\s}Z_{\rho\s}$.
As shown in the subsection 3.1.1 the $k$-deformed \sM \ for $k>0$ is
isomorphic to  $\SO(3,1)\oplus$\OSp$\oplus U(1)$; thus
there are $6=2+3+1$ Casimir operators.
In the contracted algebra four of the Casimir operators are the supersymmetrized counterparts of  \bref{4Casimir} given in \cite{Bonanos:2009wy}.
In addition  there is one $U(1)$ Casimir $\CB$ and one related to the 6th order  \OSp  Casimir operator.  

\subsubsection{ Casimir operators of \SO(N) algebras}

For $\SO(N)$ (with indefinite metric $\h=\pm1$) with generators $M_{AB}=\h_{AC}{M^C}_B=-M_{BA}$, the $n$-th order traces
\be
C^{(n)}=\frac12 tr(M^n)=\frac12 {M^{A_1}}_{A_2}{M^{A_2}}_{A_3}...{M^{A_n}}_{A_1} \ee
commute with every generator and are therefore  Casimir operators
for even $n$.  (For odd $n, \; C^{(n)}=0$ identically.)
There are $[\frac{N}2]$ independent Casimir operators since the $C^{(n)}$ for $n>N$ are
expressed as polynomial functions of lower ones using the  Cayley-Hamilton theorem.

For $\SO(3,1)$ there are two independent Casimir operators,
\bea
C_J^{(2)}&=&\frac12 {\CJ^{\mu}}_{{\nu}}{\CJ^{\nu}}_{\mu} 
=-\frac12{\CJ_{{\mu}{\nu}}\CJ^{{\mu}{\nu}}},
\nn\\
C_J^{(4)}&=&\frac12 tr(\CJ^4)=(C_J^{(2)})^2-2(\ep^{{\mu}{\nu}\rho\s}{\CJ_{{\mu}{\nu}}}{\CJ_{\rho\s}})^2.
\eea
The independent Casimir operators of $\SO(3,1)$ are thus
\bea
C_J^1&=&\frac12\CJ_{{\mu}{\nu}}\CJ^{{\mu}{\nu}}=\frac12(\CJ\CJ),\qquad
C_J^2=\frac14\ep^{{\mu}{\nu}\rho\s}\CJ_{{\mu}{\nu}}\CJ_{\rho\s}=\frac12(\CJ\7\CJ).
\label{Casso31}\eea

\vs

\subsubsection{\OSp~Casimir operators}

There are $[\frac{n}2]+[\frac{m}2]$ Casimir operators in the superalgebra $OSp(m|n)$ \cite{Jarvis:1978bc}.
 The generators of $OSp(2|4)$ are expressed as a graded antisymmetric
$OSp(2|4)$ supermatrix $M_{AB}$ with the graded symmetric $OSp(2|4)$ metric $\h^{AB}$,   ($A=(i,\A);\;i=1,2,\;\A=1,2,3,4$),
\be
M_{AB}=\pmatrix{ 0&\CB_5& \CQ_{1\B} \cr -\CB_5&0&\CQ_{2\B} \cr
\CQ_{1\A} &\CQ_{2\A}& M_{\A\B}},\qquad \h^{AB}=\pmatrix{ 1&0&0 \cr0&1&0 \cr 0 &0& -i(C^{-1})^{\A\B}},
\label{Mparam}
\end{equation}\be
{M^{A}}_{B}=\h^{AC}M_{CB}=\pmatrix{ 0&\CB_5& \CQ_{1\B} \cr -\CB_5&0&\CQ_{2\B} \cr
-iC^{-1\A\gam}\CQ_{1\gam } &-iC^{-1\A\gam}\CQ_{2\gam }& -iC^{-1\A\gam}M_{\gam\B}}.
\label{Mparam2}
\end{equation}
The 10 symmetric Sp(4) generators $M_{\A\B}$ are expressed in terms of the 10
antisymmetric \SO(3,2) generators $\CM_{{\8\mu}{\8\nu}}$ as
\be
M_{\A\B}=-\frac{i}2(C\Gam^{{\8\mu}{\8\nu}})_{\A\B}
\CM_{{\8\mu}{\8\nu}}.\label{Mab}\ee
The  $OSp(2|4)$ algebra is expressed as
\begin{eqnarray}
  \left[ M_{AB}, M_{CD}\right] _{\pm}&=&{(-i)}\left(  M_{AD}\eta
  _{CB}(-)^{C} +(-)^{(A+1)(B+1)+C}M_{BD}\eta
  _{CA}\right.\nonumber\\
  &&\left.+M_{AC}\eta_{DB}(-)^{CD+C+1} +(-)^{(AB+A+B+CD+C)}M_{BC}\eta_{DA}\right),\label{MMgraded}
\end{eqnarray}
where $(-)^{A}=+1$ for even (\SO(2)) indices $A$ and $(-)^{A}=-1$ for odd
(Sp(4)) ones.

As in the \SO(N) case we construct
 the Casimir operators of $OSp(2|4)$ using its supertraces 
\be
C^{(n)}=-\frac12\,str(M^n)=-\frac12\,(-1)^{A_1}{M^{A_1}}_{A_2}{M^{A_2}}_{A_3}...{M^{A_n}}_{A_1}.
\label{Casosp}\ee
For odd $n$ they vanish identically due to graded antisymmetry of  $M$.
There are thus three independent Casimir operators $C^{(2)},C^{(4)},C^{(6)}$. Higher power Casimir operators are not independent but rational functions of these three as follows from the super Cayley-Hamilton theorem \cite{Bonanos:2010hc}.

The explicit forms of these Casimir operators are
\bea
C^{(2)}&=&-\frac12\;str({M^2})=
\CB_5^2-{(\CM^2)}+(i\,\CQ_{i}{C^{-1}}\CQ_{i}). 
\label{C2Cas}\\
C^{(4)}&=&-\frac12\;str({M^4})=-(C^{(2)})^2+\7C_4,
\nn\\
\7C_4&=& 
-2\,\CB_5^2(\CM^2)+\frac32(\CM^2)^2+2 (\CM\7\CM)^2
\nn\\&+&
\CB_5\,\CM_{{\8\mu}{\8\nu}}(\ep^{ij}\,i\,\CQ_i\Gam^{{\8\mu}{\8\nu}}C^{-1}\CQ_j)-\frac18
(\ep^{ij}\,i\,\CQ_i\Gam^{{\8\mu}{\8\nu}}C^{-1}\CQ_j)^2
\nn\\&-&3(\CM^2)(i\,\CQ_iC^{-1}\CQ_i)
+2 (\CM\7\CM)^{\8\mu} (i\,\CQ_i\Gam_{\8\mu} C^{-1}\CQ_i)
\label{C4Cas}\\
C^{(6)}&=&-\frac12\;str({M^6}
)=-\frac12\, (C^{(2)})^3  -\frac32\, (C^{(2)})(C^{(4)})+\7C_6
\nn\\ \7C_6&=&
\CB_5^2\left(\frac94 (\CM^2)^2+3(\CM\7\CM)^2 \right)
-\frac32(\CM^2)^3-6(\CM^2)(\CM\7\CM)^2
\nn\\&+&\CB_5\left(-3 (\CM^3)^{{\8\mu}{\8\nu}}
-\frac{15}4(\CM^2)\CM^{{\8\mu}{\8\nu}}\right)
(\ep^{ij}\,i\,\CQ_i\Gam_{{\8\mu}{\8\nu}}C^{-1}\CQ_j)
\nn\\&-&
6 (\CM^2)(\CM\7\CM)^{{\8\mu}}(i\,\CQ_i\Gam_{\8\mu} C^{-1}\CQ_i)
+\left(\frac92 (\CM^2)^2 +6(\CM\7\CM)^2 \right)(i\,\CQ_iC^{-1}\CQ_i)
\nn\\&+&
\frac14(\CM^2)^{{\8\mu}{\8\nu}}(\ep^{ij}\,i\,\CQ_i\Gam_{{\8\mu}\8\rho}C^{-1}\CQ_j)
\h^{\8\rho\8\s}(\ep^{ij}\,i\,\CQ_i\Gam_{\8\s{\8\nu}}C^{-1}\CQ_j)
\nn\\&+&\frac{21}4 (\CM\7\CM)^{{\8\mu}} (i\,\CQ_i\Gam_{\8\mu} C^{-1}\CQ_i)
(i\,\CQ_iC^{-1}\CQ_i)
\nn\\&-&\frac{11}{16}(\CM^2)\left( 5(i\,\CQ_iC^{-1}\CQ_i)^2+
(i\,\CQ_i\Gam_{\8\mu} C^{-1}\CQ_i)^2 \right),
\label{C6Cas}\eea
where  $\7C_4,\, \7C_6$ are defined by subtracting terms  with higher power of
$\CB_5$ and
\bea
(\CM^2)&=&(\CM^{{\8\mu}{\8\nu}}\CM_{{\8\mu}{\8\nu}}),\quad
(\CM\7\CM)^{\8\mu} =\frac14\ep^{{\8\mu}{\8\nu}\8\rho\8\s\8\lam}\CM_{{\8\nu}\8\rho}\CM_{\8\s\8\lam},
\nn\\
(\CM^2)^{{\8\mu}{\8\nu}}&=&\CM^{{\8\mu}\8\rho}{\CM_{\8\rho}}^{\8\nu}
,\quad (\CM^3)^{{\8\mu}{\8\nu}}=\CM^{{\8\mu}\8\rho}\CM_{\8\rho\8\s}\CM^{\8\s{\8\nu}}.
\eea
These expressions are not unique due to the  identity
\be
10 (i\,\CQ_iC^{-1}\CQ_i) ^2+2 (i\,\CQ_i\Gam_{\8\mu} C^{-1}\CQ_i)^2
+(\ep^{ij}\,i\,\CQ_i\Gam_{{\8\mu}{\8\nu}}C^{-1}\CQ_j)^2=0,
\label{47}\ee
which is proved using completeness of the \SO(3,2) gamma matrices.

%%%%%%%%%%%%%%%%%%%%%%%%%%%%%%%%%%%%%%%%%%%%%%%%%
\subsubsection{Casimir operators of the \sM \ by contraction}

We have shown that
the $k$-deformed \sM \ for $k>0$ is $\SO(3,1) \oplus OSp(2|4)\oplus U(1)_\CB$
and the \sM \  is obtained by the contraction $k\to 0 \;(R\to\infty)$.
We have obtained six Casimir operators of the $k$-deformed \sM:
 two of $\SO(3,1)$ in \bref{Casso31}, three of $OSp(2|4)$ in \bref{C2Cas}-\bref{C6Cas} and $\CB$ of
$U(1)$. The Casimir operators of the \sM \  can be found by contraction
of those in the $k$-deformed algebra.

Using the relations of generators \bref{relgenSM3Rplus} we first rewrite the
Casimir operators of $U(1)_\CB$ and  $\SO(3,1)$ as
\bea
C_B&=&\CB=\frac1k B,
\\
C_J^1&=&\frac12(\CJ\CJ)=\frac12(M_{{\mu\nu}}+\frac{Z_{{\mu\nu}}}{k})(M^{{\mu\nu}}+\frac{Z^{{\mu\nu}}}{k}),
\\
\nn\\
C_J^2&=&\frac14\ep^{{\mu\nu}\rho\s}\CJ_{{\mu\nu}}\CJ_{\rho\s}=\frac14\ep^{{\mu\nu}\rho\s}(M_{{\mu\nu}}+\frac{Z_{{\mu\nu}}}{k})(M_{\rho\s}+\frac{Z_{\rho\s}}{k}).\label{CasmirJ2C}
\eea
In the contraction the leading terms 
{give} the Casimir operators of the contracted algebra. In 
{this way} we get three Casimir operators of \sM \ as
\bea
\CC_B&=&\lim_{k\to0}k\, C_B\;=\;B,
\\
\CC_J^1&=&\lim_{k\to0}k^2\, C_J^1\;=\;\frac1{2}(ZZ),
\nn\\
\CC_J^2&=&\lim_{k\to0}k^2\, C_J^2\;=\;\frac1{2}(Z\7Z).
\label{CasmirSM1}
\eea
For  the \OSp Casimir operator
\bea
C^{(2)}&=&
\CB_5^2-{(\CM\CM)}+(i\,\CQ_{i}{C^{-1}}\CQ_{i})=
\frac1{k^2}\left(B^2- (ZZ)\right)+\frac1{k}\left(\CC_2\right)+...
\label{C2Cas1}\eea
The leading order term $(\CC_B)^2 -2 (\CC_J^1)$ is a function of the Casimir operators, which we have found above.
In this case the independent Casimir appears as the next leading term
\cite{Casalbuoni:2008ez}.
If we consider the combination in which the leading $k^{-2}$ terms cancel
\bea
\CC_2&=&\lim_{k\to0}k\, \left(C_2+2 (C_J^1)-{(C_B)^2}\right)
=\;2 \left(P^2+(MZ)+{i}\bS C^{-1}\bQ-\,B_5\,B\right)
\label{SMCas123C}
\eea
we obtain the mass Casimir of the \sM \ found in \cite{Bonanos:2009wy}. Note that sub-leading term $(MZ)$ in $(C_J^1)$ also contributes to the $\CC_2$ term.
\vs

In the same way 
the leading $k^{-4}$ terms in $\7C_4$ are cancelled by a
combination of other Casimir operators
\bea
\7C_4-6 (C_J^1)^2+2 (C_J^2)^2+4\,{(C_B)^2}(C_J^1)&=&
\frac1{k^3}\left(\CC_4\right)+ \frac1{k^2}\left(...\right)+
\eea
leaving
\bea
\CC_4&=&\lim_{k\to0}k^3\,\left(\7C_4-6 (C_J^1)^2+2 (C_J^2)^2+4\,{(C_B)^2}(C_J^1)
\right)+4\,(\CC_J^1)(\CC_2)  
\nn\\ 
&\=&8(P\7Z)^2+2(Z\7Z)(M\7Z)
+(4\,{B}^2\,-2\,(ZZ)\,)\left(
P^2+(MZ)\right)
\nn\\
&-&2\,(ZZ) \left({i}\bS C^{-1}\bQ 
\right)\,
-2\,{B}\,Z^{\mu\nu}(i\,\bS\,\G_{\mu\nu}\G_5\,C^{-1}\,\bQ)-2(Z\7Z)\,
(i\,\bS\,\G_5\,C^{-1}\,\bQ).
\nn\\
&+&\left(8(P\7Z)^\mu\,+4\,{B}\,P^{\mu}\right)
({i}\,\bS\,\G_{\mu}\G_5\,C^{-1}\,\bS)\,
-4\,(i\,\bS\,\G_5\,C^{-1}\,\bS)^2 ,
\label{SMCas4C0}
\eea
where $(P\7Z)^\mu=P_\nu\7Z^{\nu\mu}$.
Finally, the leading $k^{-6}$ terms in $\7C_6$ are cancelled by the following
combination of other Casimir operators
\bea
\CC_6&=&\lim_{k\to0}k^5\,\left(\7C_6+12(C_J^1)^3-12(C_J^1)(C_J^2)^2-9({C_B})^2(C_J^1)^2+3 (C_B)^2(C_J^2)^2\right)
\nn\\&&-
\left(9\,(\CC_J^1)^2-3(\CC_J^2)^2\right)\,(\CC_2)
\nn\\
&=&\left(\frac92\,(ZZ)^2 -9 B^2(ZZ)-
\frac32\,(Z\7Z)^2 \right)((PP)+(MZ))
\nn\\&+&
12(B^2-2(ZZ))\left((P\7Z)^2+\frac14(M\7Z)(Z\7Z)\right)
\nn\\&+&6(ZZ) (Z\7Z) \, (i\,\bS\,\G_5 C^{-1}\,\bQ)
+\frac32\,B\left(3(ZZ)\,Z^{\mu\nu} -(Z\7Z)\7Z^{\mu\nu} \right)
 (i\,\bS\G_{\mu\nu}\G_5\,C^{-1}\,\bQ)
\nn\\&-&3\,B\left(5(ZZ)\,P^\mu+4 (PZZ)^\mu\right) (i\,\bS\,\G_{\mu}\G_5\,C^{-1}\,\bS) -24\,(ZZ)\,(P\7Z)^\mu (i\,\bS\,\G_{\mu}\G_5\,C^{-1}\,\bS)
\nn\\&+&
\frac32\,(3(ZZ)^2-(Z\7Z)^2)   \, (i\,\bS\,C^{-1}\,\bQ)
+12\,(ZZ)\,(i\,\bS\,\G_5\,C^{-1}\,\bS)^2,
\eea
where
$
(PZZ)^\mu =P^\nu Z_{\nu\rho}Z^{\rho\mu}.
$
When all fermions vanish, $\CC_6$ becomes a function of lower order Casimir operators.
This explains why $\CC_6$ does not appear in the bosonic Maxwell algebra.

In summary there are 6 Casimir operators in the \sM, % (nb0125OSpPB.nb)
$
\CC_B,\, \CC_J^1,\,  \CC_J^2,\,  \CC_2,$ $\,  \CC_4,\, \CC_6.
$
Note that in the defining equations there appear products of generators, which are not (anti)commuting.
A careful investigation shows that we obtain the Hermitian Casimir
operators if we use the operator ordering presented above.
 \vs

%%%%%%%%%%%%%%%%%%%%%%%%%%%%%%%%%%%%%%%%
\subsection{Casimir operators of D=3 \sMs}

In three dimensions the \sM \ is also obtained from the deformed algebras by contraction. Similarly as in four dimensions we obtain the Casimir operators from those of the deformed algebra by contraction.
The $k^+$-deformed algebra is the direct sum of
$\SO(2,1)\oplus OSp(1|2)\oplus OSp(1|2)$ and its Casimir operators are
\be
C_\CJ= \CJ_\mu \CJ^\mu
\ee
for the $\SO(2,1)$ and
\be
C^\pm= 4\CM_\mu^\pm \CM^{\pm\mu}\,\mp\, i\,\bQ^\pm C^{-1}\bQ^\pm
\ee
from each of the $OSp(1|2)$'s. We recall that
there is only one Casimir operator in $OSp(1|2)$.

The Casimir operators of \sM \ in D=3 is obtained by contracting them. Using relation of generators dual to \bref{relkplus}
\bea
C_\CJ&=& 
R^4\,Z^2+R^2 \,2\,MZ+M^2,
\nn\\
C^\pm&=& 
R^4\,Z^2\pm R^3(2PZ-i\bS C^{-1}\bS)+R^2(P^2+i\bQ C^{-1}\bS)\mp
R (\bQ C^{-1}\bQ)
\eea
we get three Casimir operators of the \sM \  in three dimensions,
\bea
C_1&=&\lim_{R\to\infty}\frac1{R^4}C_\CJ=Z_\mu Z^\mu,\\
C_2&=&\lim_{R\to\infty}\frac1{2R^2}((C^++C^-)-2C_\CJ)=P_\mu^2-2M_\mu Z^\mu+i\bQ C^{-1}\bS,\\
C_3&=&\lim_{R\to\infty}\frac1{2R^4}(C^+-C^-)=2P_\mu Z^\mu-i\bS C^{-1}\bS.
\eea
\vs

%%%%%%%%%%%%%%%%%%%%%%%%%%%%%%%%%%%%%
\section{Conclusions and Outlook}

In this  paper we  presented all deformations of D=3 and D=4 Maxwell
superalgebras. The operations of deformation and supersymmetrization of Maxwell  algebra can be done in different order 
 resulting in different superalgebras (see Fig.1).

%%%%%%%%% FIG.1 %%%%%%%%%%%%%%%%
\bea
\matrix{{\bf Maxwell}  &&&{\rm SUSY}&\;\;&  {\bf Maxwell}\cr
{\bf algebra}& &\;\;&\longrightarrow&& {\bf superalgebra }\cr 
&&&&&|\cr
|&&&&& {\rm deformation} \cr
{\rm deformation} &&&&&\downarrow \cr
\downarrow &&&&&{\bf (a)\;deformed\;Maxwell}\cr
&&&&& {\bf superalgebra}\cr 
{\bf deformed\;Maxwell} & &&{\rm SUSY}&& \nparallel \cr
{\bf algebra}&&&\;\longrightarrow &&{\bf (b)}\;{\bf superextension\; of\; }\cr
&&&&&{\bf deformed\; Maxwell\; algebra} \cr}
\nn \eea \vs
\begin{center}
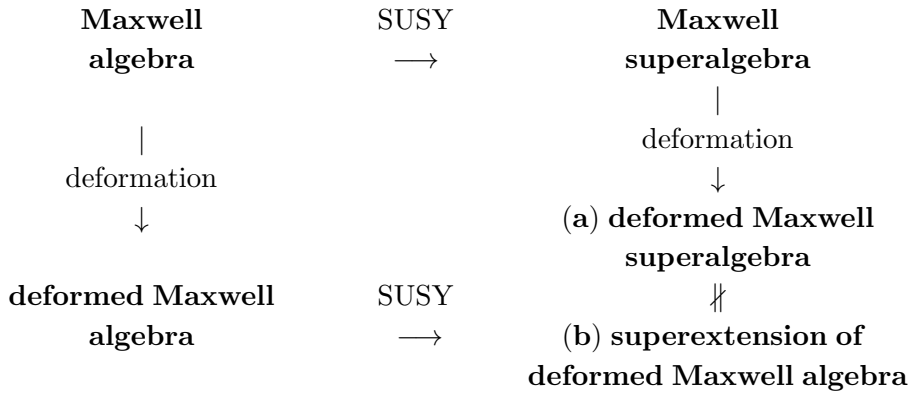
Fig. 1 ~~~  Two ways of modifying the Maxwell algebra
\end{center}
\vs

  We are following the upper path leading to (a) in Fig. 1, on the other hand  Soroka and Soroka \cite{Soroka:2010ht} recently have chosen the other path leading to (b). The results do not coincide, because
the supersymmetrization of Maxwell algebra requires eight real supercharges
while that of the deformed Maxwell algebra, $\SO(3,1)\oplus \SO(3,2)\to \SO(3,1)\oplus OSp(1|4)$, only needs four real supercharges.

It is plausible to assume, by analogy with the N=1 case considered in this paper, that the N-extended Maxwell superalgebra can be obtained by suitable contraction of the $OSp(2N|4)$ algebra\footnote{In principle one can obtain the same algebra by contraction of the N-extended D=4 dS superalgebra, but algebraically this procedure is more complicated due to the presence of quaternionic Majorana subsidiary conditions.} and it will have 8N real supercharges. We conjecture as well that the D-dimensional Maxwell superalgebra can be obtained by contraction of the D-dimensional AdS superalgebra, i.e., it exists in those dimensions (D=2,3,4,5,7) in which we can formulate AdS superalgebras\footnote{One can also obtain the D-dimensional \sM \ by contracting  D-dimensional dS superalgebras, but dS superalgebras exist only for D=2,3,4,5 (see also Appendix B).} \cite{vanHolten:1982mx}. An interesting question is whether it is possible to construct \sMs \ with bosonic Maxwell subalgebra in other dimensions, e.g., in D=10 and D=11.

Our considerations above are only valid if we assume that the supersymmetrization of Maxwell algebra, which can be called {\it minimal},  satisfies the following two assumptions:
\begin{enumerate}
\item it is an enlargement of Poincar{\`e} superalgebra with the four-momenta present in the anticommutator of supercharges, 
\item its bosonic sector is the direct sum of Maxwell algebra \bref{max}-\bref{max1} and possibly some internal sector.
\end{enumerate}
For if we allow generalized Maxwell superalgebras,
 for example, with the following basic SUSY relation  $\{Q,Q\} = bZ$  (see \cite{Soroka:2004fj}) that contradicts our first assumption,  or with a bosonic  sector that is a non-Abelian {\it enlargement} of Maxwell algebra, we open up a ``Pandora's box'' of possibilities.
The simplest Abelian  enlargement consists in adding to Maxwell algebra some central charges, like $B$ and $B_5$ in D=4 (see section 2), but one can also enlarge Maxwell algebra
into a semisimple group with additional bosonic generators. One category of such extensions is described by the Bergshoeff-Sezgin superalgebra \cite{Bergshoeff:1995hm}. If we consider D=10, 11, which play a crucial role in string/M theory, one can look for contractions of generalized D=10 and D=11 AdS algebra $OSp(1|32)$, see for example \cite{Bergshoeff:2000qu}, and look for the corresponding 
generalized D=10 and D=11 \sMs.

%%%%%%%%%%%%%%%%%%%%%%%%%%%%
\vskip 6mm

 \appendix
{\bf \Large Appendix }
\section{Notations and conventions}

Here we summarize our notations and conventions.

We use a positive signature space-time metric $\h_{\mu\nu}=diag(-,+,...,+)$.
Antisymmetrization is defined by
\be A_{[\mu}B_{\nu]}=A_\mu B_\nu-A_\nu B_\mu.\ee
Our convention of hermitian conjugation for odd generators  is
$\bL^\dagger=\bL,\;\bQ^\dagger=-\bQ,$ so that $\bQ_\A\bL^\A$ in the MC one-form $\W$ is Hermitian,
\be
(\bQ_\A\bL^\A)^\dagger={\bL^\A}^\dagger\bQ_\A^\dagger={\bL^\A}(-\bQ_\A)=(\bQ_\A\bL^\A).
\ee
The conjugate for Majorana spinors $\ba\bL$'s are defined using charge conjugation matrix
$C$ as
\be
 \ba\bL_\A=\bL^\B{C}_{\B\A}.
 \ee

In 4 dimensions we use the gamma matrices in the Majorana representation
\be
  \G_0=i\s_2\otimes I_2,\quad \G_1=\s_3\otimes \s_1,\quad
\G_2=\s_3\otimes\s_3,\quad \G_3=\s_1\otimes I_2.
\label{4Dgamma}\ee
They are all real and verify  the Clifford algebra
\be
\{\G_{ \mu},\G_{ \nu}\}=2\,\h_{{ \mu}{ \nu}},\quad \h_{{ \mu}{ \nu}}=(-,+++),\quad { \mu},{ \nu}=0,1,2,3.
\ee
The $\SO(3,1)$ Lorentz generators are
$
\G_{\mu\nu}=\frac12[\G_{\mu},\G_{\nu}]=-\G_{\nu\mu}$
 and satisfy
\be
[\G_{{ \mu}{ \nu}},\G_{ \rho \s}]=2\left(
\,\h_{ \rho{[ \nu}}\G_{{ \mu]} \s}-\h_{ \s{[ \nu}}\G_{{ \mu]} \rho}\right).
\ee
The charge conjugation matrix $C$ and $\gamma_5$ are also real
 \be
  C=\G_0,\quad \G_5=\G_{0123}=\s_1\otimes i\s_2
\ee
and satisfy
\be
C^T=-C=C^{-1},\quad \G_\mu^T=C\G_\mu C^{-1}
\ee
then
\be
(C\G_\mu)^T=-(C\G_\mu),\qquad (C\G_{\mu\nu})^T=(C\G_{\mu\nu}),\qquad
(C\G_5)^T=-(C\G_5).
\ee
The cyclic identities necessary for the closure of the superalgebras are
\be
(C\G_\mu)_{(\A\B}(C\G^\mu)_{\gam\D)}=0,\quad (\A\B\gam\D)\; {\rm symmetric\; sum}.
\ee
\vs

The $\SO(3,2)$ gamma matrices $\Gam_{\8\mu}$'s are expressed by using the Majorana representation of the $\SO(3,1)$ gamma matrices \bref{4Dgamma} as
\be
\Gam_\mu=\G_\mu\G_5,\quad
\Gam_4=-\,\G_5, \label{SO32GamMat}
\ee
and verify the $\SO(3,2)$ Clifford algebra
\be
\{\Gam_{\8\mu},\Gam_{\8\nu}\}=2\,\h_{{\8\mu}{\8\nu}},\quad \h_{{\8\mu}{\8\nu}}=(-,+++;-),\quad {\8\mu},{\8\nu}=0,1,2,3,4.
\label{Clif32}\ee
The $\SO(3,2)$ generators $\Gam_{{\8\mu}{\8\nu}}$ are
\be
\Gam_{\mu\nu}=\frac12[\Gam_{\mu},\Gam_{\nu}]=\G_{\mu\nu},\quad
\Gam_{\mu 4}=\Gam_{\mu}\Gam_{4}=\,\G_\mu,
\ee
and satisfy
\be
[\Gam_{{\8\mu}{\8\nu}},\Gam_{\8\rho\8\s}]=2\left(
\,\h_{\8\rho{[\8\nu}}\Gam_{{\8\mu]}\8\s}-\h_{\8\s{[\8\nu}}\Gam_{{\8\mu]}\8\rho}\right).
\ee
It gives the real Majorana representation of the algebra 
\bref{Clif32} with the
charge conjugation matrix $C=\G_0=-\Gam_0\Gam_4$,
\be
(C)^T=-(C),\quad(C\Gam_{\8\mu})^T=-(C\Gam_{\8\mu}),\quad
(C\Gam_{{\8\mu}{\8\nu}})^T=(C\Gam_{{\8\mu}{\8\nu}}).\label{TranposeRels}
\ee
\vs

The $\SO(4,1)$ gamma matrices $\Gam_{\8\mu}$ are
\be
\Gam_\mu=\G_\mu\G_5,\quad
\Gam_4=i\,\G_5. \label{SO41GamMat}
\ee
They verify the $\SO(4,1)$ Clifford algebra
\be
\{\Gam_{\8\mu},\Gam_{\8\nu}\}=2\h_{{\8\mu}{\8\nu}},\quad \h_{{\8\mu}{\8\nu}}=(-,+++;+),\quad {\8\mu},{\8\nu}=0,1,2,3,4.
\ee
 The  $\SO(4,1)$ generators $\Gam_{{\8\mu}{\8\nu}}$ are
\be
\Gam_{\mu\nu}=\frac12[\Gam_{\mu},\Gam_{\nu}]=\G_{\mu\nu},\quad
\Gam_{\mu 4}=\Gam_{\mu}\Gam_{4}=-i\,\G_\mu,
\ee
where 
\be
[\Gam_{{\8\mu}{\8\nu}},\Gam_{\8\rho\8\s}]=2\left(
\,\h_{\8\rho{[\8\nu}}\Gam_{{\8\mu]}\8\s}-\h_{\8\s{[\8\nu}}\Gam_{{\8\mu]}\8\rho}\right).
\ee
Using the same charge conjugation matrix $C=\G_0=i\Gam_0\Gam_4$,  we find that the same relations \bref{TranposeRels} hold also for $\SO(4,1)$ gamma matrices.
In contrast to the AdS case these gamma matrices are complex but satisfy
\be
\Gam_{\8\mu}^*=-\Gam_4\,\Gam_{\8\mu}\,\Gam_4,\qquad
\Gam_{{\8\mu}{\8\nu}}^*=\Gam_4\,\Gam_{{\8\mu}{\8\nu}}\,\Gam_4.
\label{stargamma2}\ee
\vs

%%%%%%%%%%%%%%%%%%%%%%%%%%%%%%%%%%%%%%%%%%%%%%%%%%%%%%%%%%%%%%%%%%%%%%5
\section{Quaternionic (super)groups and (super)algebras}
\subsection{Quaternionic algebras}
Quaternions $q$, the elements of quaternionic space $H^n=(q_1...q_n)$, are written using
\be
q=q^{(0)}+e_r\,q^{(r)}\in H, \qquad q^{(0)},q^{(r)}\in R,
\ee
where the quaternionic units $e_r$'s satisfy the multiplication rule
\be
e_r\,e_s=-\delta_{rs}+\epsilon_{rst}\,e_t.\label{BQalgebra}
\ee
Quaternionic conjugate   $\ba q$ of  $q$ is defined by $\ba{q}=q^{(0)}-e_r\,q^{(r)}$.
Remember $ q-\ba q=0, \to q^{(r)}=0,\; (r=1,2,3),
\; q+\ba q=0,\to q^{(0)}=0$ and it holds $ \ba{p\,q}=\ba q \;\ba p$.

\subsection{Quaternionic groups and corresponding  Lie algebras \cite{Gilmore:1974}
\cite{Tits:1967}}
{\underline {i) The pseudounitary quaternionic (unitary) group $U(n-k,k;H)$;}}
\vs
It  consists of  the generators of linear quaternionic $n \times n$ matrix transformations preserving  the Hermitian scalar product
\be
\braket{\ba{q}}{p}=\sum_{i=1}^{n-k} \ba{q}_i\, p_i-\sum_{j=1}^k  \ba{q}_{n-k+j}\, p_{n-k+j} \equiv \sum_{K,L=1}^n \ba{q}_K \eta_{KL}p_L. \label{BQnorm}
\ee
The transformation of $q$ and $p$ is
\bea
p_K \too U_{KL}p_L,\qquad q_K \too U_{KL}q_L,\qquad
\ba q_K \too  \ba q_L \ba{U_{KL}}\equiv \ba q_L U^\dagger_{LK}.
\eea
where $\dagger$ is defined using quaternionic conjugation. Then the condition of invariance of the metric is
\be
U^\dagger\,\h\,U=\h,\quad\too \quad\h_{KN}\ba{U_{KL}}U_{NM}=\h_{LM}.
\label{BUcon}\ee
Number of independent conditions (in real components) following from \bref{BUcon}
are $ 4\; \frac{n(n-1)}{2}$ for off-diagonal and $n$ for diagonal components, i.e.,
the  number of independent real group parameters  is
$4n^2-\,4\, \frac{n(n-1)}{2}-n\,=n(2n+1).$
For arbitrary $n,\,k$ the norm \bref{BQnorm} defines the pseudounitary quaternionic (unitary) group $U(n-k,k;H)$.

\vs
{\underline {ii) The antiunitary quaternionic group $U_\alpha(n;H)$ }}
\vs
One can define as well in quaternionic space 
the antiunitary quaternionic group $U_\alpha(n;H)$ as preserving the antihermitian quaternionic scalar product \cite{Tits:1967}\cite{Kugo:1982bn}
\be
\braket{\ba{q}}{p}_A=\sum_{K,L=1}^n \ba{q}_K A_{KL}p_L. \label{BQAnorm}
\ee
with
\be
A_{KL}=-\ba{A}_{LK},\qquad\too\qquad A=-A^\dagger.
\ee
For even $n$ ($n=2k$) one can choose $A_{KL}$ to be  a real antisymmetric $2k \times 2k$ matrix, and for odd $n=2k+1$ the last element  $A_{2k+1,2k+1}$ should be purely quaternionic (one usually assumes that it is given by the unit $e_2$).
Then the condition of invariance of the metric is
\be
U^\dagger\,A\,U=A.
\label{BUacon}\ee
The number of independent conditions (in real components) following from \bref{BUacon} is
$4\, \frac{n(n-1)}{2} $ for off-diagonal and $3n$ for diagonal components, i.e.,
the  number of independent real group parameters  is
$4n^2-\,4\, \frac{n(n-1)}{2}-3\,n\,=n(2n-1).$
For arbitrary $n$ the norm \bref{BQAnorm} defines the  antiunitary quaternionic group $U_\alpha(n;H)$.

\vs

We could also consider orthogonal quaternionic groups $O(n;H)$ and
symplectic ones $Sp(2n;H)$.
These two quaternionic groups
are defined as preserving the symmetric and antisymmetric scalar products respectively,
\be
\braket{{q}}{p}_S=\sum_{K,L=1}^n {q}_K^T \h^{KL}p_L,\qquad
\braket{{q}}{p}_A=\sum_{K,L=1}^n {q}_K^T A^{KL}p_L, \label{OSnorm}
\ee
where the transposed quaternions are described by partial
inversion $q\to q^T = q^{(0)} + e_3 q^{(3)} -e_2 q^{(2)} +e_1 q^{(1)}
$\footnote{The way the transposed quaternions are defined follows from
the complex realization \bref{BQunitToPauli}. It should be recalled that the standard
framework for the description of all Lie algebras are complex algebras,
and the quaternionic variables are rather used as auxiliary notation.
}.
However  these Lie algebras are related to the unitary and antiunitary ones by
the following relations,
\bea
O(n;H)&=& U_\alpha (n;H), \qquad \qquad n(2n-1)\qquad \mbox{real parameters} \nn \\
Sp(2n;H)&=& U ({n},{n};H), \quad\qquad 2n(4n+1)\qquad \mbox{real parameters}\label{BUUparameters}
\eea
as can be shown by using the complex realization of the  quaternionic scalar products \bref{OSnorm}.
\vs

{\underline {iii) The complex representation of quaternions and quaternionic groups }}
\vs

In order to describe the quaternionic groups and algebras using complex variables, one can introduce the known representation of the quaternionic algebra \bref{BQalgebra} in terms of 2$ \times$2 Pauli matrices
\be
e_r \leftrightarrow -i\,{(\s_r)}_{ij},\quad \to\quad
q=\pmatrix{q^{(0)}-iq^{(3)}&-q^{(2)}-iq^{(1)}  \cr
           q^{(2)}-iq^{(1)} &q^{(0)}+iq^{(3)}}=\pmatrix{z^1&-z^{2*}\cr z^2&z^{1*}} . \label{BQunitToPauli}
\ee
Multiplication of quaternions is the product of their matrix representations.
The conjugate $\ba q$ is  described as the hermitian conjugate of the complex matrix $q$\be
\ba q= q^{(0)}-i\,{(\s_r)}(-q^{(r)}) ,\quad \to\quad
\ba q=\pmatrix{q^{(0)}+iq^{(3)}&q^{(2)}+iq^{(1)}  \cr
           -q^{(2)}+iq^{(1)} &q^{(0)}-iq^{(3)}}
=\pmatrix{z^{1*}&z^{2*}\cr- z^2&z^{1}} . \label{BQunitToPauli2}
\ee
In this way we identify a quaternion with a pair of complex numbers,
\be
q\leftrightarrow \pmatrix{z^1\cr z^2},\qquad q\in H,\qquad
\pmatrix{z^1\cr z^2}\in C^2.
\ee
The components of the second column of $q$ are determined from those of the first.

Using the substitutions \bref{BQunitToPauli},  \bref{BQunitToPauli2}, one can express the quaternionic scalar product \bref{BQnorm} as pairs of complex numbers, namely\footnote{
The complex variables $u^k$ are related to the components of $p$ as $(z^1,z^2)$ are to those of $q$ -- see \bref{BQunitToPauli}.}\cite{Gilmore:1974}\cite{Lukierski:1983qc}
\bea \braket{\ba{q}}{p}_S&=&\pmatrix{{z}_A^{1*}\eta_{AB}u_B^1+{z}_A^{2*}
\eta_{AB}u_B^2 \cr z_A^1\eta_{AB}u_B^2-z_A^2\eta_{AB}u_B^1}=
\pmatrix{({z}_A^{1*},{z}_A^{2*})\pmatrix{\eta_{AB}&0\cr0&\eta_{AB}}
\pmatrix{u_B^1\cr u_B^2} \cr (z_A^1,z_A^2)\pmatrix{0&\eta_{AB}\cr -\eta_{AB}&0}\pmatrix{u_B^1\cr u_B^2}}.
\label{B39}\eea
Invariant transformations of the inner product are those of the $2n$ complex
vectors that keep the two components of \bref{B39} invariant. It follows that in complex $2n$ dimensional space $(z^1_A,z^2_A)\in C^{2n}$
one can introduce a pair of complex scalar products; first describing
$U(2n-2k,2k;C)$
with diagonal pseudo-metric $\7\h=\pmatrix{\h_{AB}&0\cr 0&\h_{AB}}$
and second with $2n\times 2n$ antisymmetric metric $\7A=\pmatrix{0&\h_{AB}\cr-\h_{AB}&0}$
describing the group $Sp(2n;C)$.
One obtains in complex notation the following equivalence as follows
\bea
U(n-k,k;H)&=&U(2n-2k,2k;C)\cap Sp(2n;C)=USp(2n-2k,2k;C).
\eea

For the  antihermitian quaternionic inner product \bref{BQAnorm} the complex components are written as
\bea
\braket{\ba{q}}{p}_A&=&\pmatrix{{z}_A^{1*}A_{AB}u_B^1+{z}_A^{2*}A_{AB}u_B^2
\cr{z}_A^1A_{AB}u_B^2-{z}_A^2A_{AB}u_B^1}=
\pmatrix{({z}_A^{1*},{z}_A^{2*})\pmatrix{A_{AB}&0\cr0&A_{AB}}
\pmatrix{u_B^1\cr u_B^2} \cr (z_A^1,z_A^2)\pmatrix{0&A_{AB}\cr -A_{AB}&0}\pmatrix{u_B^1\cr u_B^2}}.\qquad
\label{B40}\eea
Here  the symmetric  $2n\times 2n$ matrix $\8\h=\pmatrix{0&A_{AB}\cr-A_{AB}&0}$ has the signature $(1,...,1,-1,...,-1)$ and defines the complex group $O(n,n;C)=O(2n;C)$\footnote{The orthogonal complex algebras do not have signatures - $O(n;C)$ is the complexification of real algebras $O(n-k,k)$ for $k=0,1...n$.
} and the antisymmetric $\8A=\pmatrix{A_{AB}&0\cr 0&A_{AB}}$ defines the antiunitary complex group  $Sp(2n;C)$ which is equivalent with $U(n,n;C)$,
\cite{Tits:1967}\cite{Hasiewicz:1983zh}.
Then we obtain
\bea
U_\alpha(n;H)&=& O(2n;C)\cap U(n,n;C)=O^*(2n).
\eea

%%%%%%%%%%%%%%%%%%%%%%%%%%55
\subsection{Quaternionic supergroups and applications to space-time symmetries}

There exists only one infinite series of norm-preserving supergroups
$UU_\A(n-k,k|m;H)$ which leave invariant the following graded metric in quaternionic superspace
$H^{n|m}$ spanned by the graded quaternionic vectors $(q_1\dots q_n; \T_1\dots \T_m)$
\be
\braket{{\cal Q}}{{\cal P}}=\sum_{K,L=1}^n \ba{q}_K \eta_{KL}p_L+\sum_{\A,\beta=1}^m \ba{\T}_\A A_{\A\beta}\T_\beta. \label{BsuperNorm}
\ee
The variables $\T_\A=\T_\A^{(0)}+\T_\A^{(r)}e_r$ are
quaternions with Grassmann-valued components, i.e.,
$\T_\A^{(0)},\,\T_\A^{(r)}$ form the basis of a real
$4m$-dimensional Grassmann algebra. The bosonic sector of the supergroup $UU_\A(n-k,k|m;H)$ is given by the product of quaternionic groups
$U(n-k,k;H)\otimes U_\A(m;H)$.

The quaternionic graded scalar product \bref{BsuperNorm} can be expressed equivalently by a pair of complex  graded scalar products defining the pair of complex supergroups $U(2n-2k,2k|m,m)$ and $OSp(2m|2n;C)$. We have the following equivalence in quaternionic and complex notation \cite{Kac:1977qb}\cite{Lukierski:1983qc}
\be
UU_\A(n-k,k|m;H)  = U(2n-2k,2k|m,m)\cap OSp(2m|2n;C).
\ee

In the description of space-time symmetries the quaternionic groups 
are given as
\be
{\ov{{O(4,1)}}}=U(1,1;H)=USp(2,2)
\ee
and for D=7 anti-de-Sitter symmetries \cite{Gilmore:1974} as
\be
{\ov{{O(6,2)}}}=U_\A(4;H)=O^*(8).
\ee
The $n$-extended D=4 dS supergroup is \cite{Kugo:1982bn}-\cite{Pilch:1984aw}
\be
UU_\A(1,1|n;H)=U(2,2|n,n)\cap OSp(2n|4;C)
\ee
with internal symmetries $U(n;H)=USp(2n)$,
and the D=7 AdS supergroup,
after using the relation $U(n|m)=U(m|n)$ valid for graded unitary supergroups,
takes the form \cite{Kugo:1982bn}\cite{Hasiewicz:1983zh}
\be
U_\A U(4|n;H)=U(4,4|2n)\cap OSp(2n|8;C).
\ee
The  internal symmetries in D=7 are 
given by $U_\A (n;H)=O^*(2n)$.

\subsection{D=4 dS superalgebra}
The graded de-Sitter algebras in D=2,3,4,5 are the same as D=3,4,5,6 Lorentz algebras.
They are described by a symplectic series of real, complex or quaternionic Lie algebras $Sp(2,F), \, (F=R,C,H)$ and quaternionic $Sl(2,H)$\footnote{In the third relation in \bref{BD234algebras} we have used the second of \bref{BUUparameters}.}
\bea
D=2: \quad sp(2;R)&=&sl(2;R)=o(2,1)\nn \\
D=3: \quad sp(2;C)&=&sl(2;C)=o(3,1) \label{BD234algebras} \\
D=4: \quad sp(2;H)&=&u(1,1;H)=o(4,1)\nn\\
D=5: \quad sl(2;H)&=&su^*(4)=o(5,1).\nn
\eea
If we employ the relations \bref{BD234algebras} as linking various groups, we obtain  the series of double spinorial coverings of $O(D,1)$ for D=2,3,4,5:
\bea
\ov{{\SO(2,1)}}&=&SL(2;R)\nn\\
\ov{{\SO(3,1)}}&=&SL(2;C)  \label{BD234cover}\\
\ov{{\SO(4,1)}}&=&U(1,1,H)\nn\\
\ov{{\SO(5,1)}}&=&SL(2,H).\nn
\eea

The supersymmetrization of space-time symmetries requires the knowledge of the fundamental spinor representation of the corresponding spinorial covering group. We shall assume that
the (super)algebra is supersymmetrized by the introduction of supercharges, which transform under the transformations of space-time symmetry group as the fundamental spinor representation.\footnote{By this assumption we exclude so-called vector-supersymmetries \cite{Casalbuoni:2008ez}.}

The spinorial coverings \bref{BD234cover} are supersymmetrized as follows:
\bea
D=2&:& \quad OSp(N|2;R) \; \mbox{or} \; SL(2|N;R)\nn \\
D=3&:& \quad SL(2|N;C) \label{Bsuper D234cover} \\
D=4&:& \quad UU_\A(1,1|N;H)\nn\\
D=5&:& \quad SL(2|N;H).\nn
\eea

The N=1,\, D=4 dS superalgebra $UU_\A(1,1|1;H)$ used in this paper can be translated into complex notation by using the relations \bref{BQunitToPauli}, \bref{BQunitToPauli2}, but this is quite a tedious procedure (see for example the complex description of $sl(2|N;H)$ in \cite{Lukierski:1983qc}). In this paper we obtained the N=1,\, D=4 dS superalgebra (see \bref{superdS5}) by observing its correspondence with the well-known N=2,\, D=4 AdS superalgebra (see \bref{superAdS5}). Such a procedure follows from the property that $OSp(2|4;R)$ as well as $UU_\A(1,1|1;H)$ are two different real forms of $OSp(2|4;C)$ and consists of the following steps:
\begin{enumerate}
\item{Replace the real $4\times 4$ matrices  \bref{SO32GamMat} describing $O(3,2)$ Dirac matrices by the complex $4\times 4$ matrices \bref{SO41GamMat} giving the $O(4,1)$ Dirac algebra.\footnote{The complexification in our paper is obtained simply by multiplying the fifth real $O(3,2)$ gamma matrix by the imaginary unit $i$.}}
\item{The choice of  $O(4,1)$ gamma matrices dictates the choice of $O(4,1)$ conjugation matrix $C$, satisfying the defining relations \bref{TranposeRels} for $O(4,1)$ Dirac matrices.
In our paper we select the same explicit form of $C$-matrix for $\SO(3,2)$ and $\SO(4,1)$ (see appendix A).
}
\item{Replace the N=2 real $OSp(2|4)$ supercharges by complexified supercharges describing the complex-holomorphic superalgebra $OSp(2|4;C)$.}
\item Impose
 invariance of the superalgebra  $OSp(2|4;C)$ under the quaternionic Majorana conjugation implying that the supercharges are $\SO(4,1)$ quaternionic spinors
\be
\CQ_{\A}\to a\,\ep^{ij} \,{{(\Gam_4)}_\A}^\B\,\CQ^\dagger_{\B j}.\label{BQaQdagger}
\label{BB36}\ee
where  $|a|=1$ due to the involutive character of the mapping  \bref{BQaQdagger} (in \bref{symMajo2} the choice $a=i$ is made).
\end{enumerate}

Our superalgebra \bref{superdS5} satisfies all these requirements characterizing the   $UU_\A(1,1|1;H)$  superalgebra, namely
\begin{itemize}
\item{It is covariant under $U(1,1;H)\otimes U_\A(1;H)=USp(2,2)\otimes O(2)$} transformations.
\item{The 8 complex charges do satisfy the quaternionic Majorana condition \bref{symMajo2}.}
\end{itemize}

In conclusion, we would like to add that N=1,\, D=4 dS supergravity, as a gauge theory on $UU_\A(1,1|1;H)$, was constructed first in \cite{Lukierski:1984pp}.
Unfortunately, it has been shown \cite{Pilch:1984aw}\cite{Lukierski:1984it} that such a theory necessarily contains ghost states.

\vs

\vs
 {\bf Acknowledgments}

We acknowledge discussions with Jose A. de Azcarraga, Jaume Gomis, Toine Van Proeyen. One of the authors (J.L.) would like to thank Professors Yan-Gang Miao
and Mo-Lin Ge for their superb hospitality at Nankai University during
the time of completion of this paper. We also acknowledge financial support from projects FPA2007-66665-C02-01, 2009SGR502, Polish Ministry of Science and High Education grant NN202 318534 and Consolider CPANCSD2007-00042.

\end{document}